\documentclass[epj]{svjour}
\usepackage[sort&compress,numbers]{natbib}
\usepackage{amssymb}
\usepackage{graphicx}
\usepackage{geometry}
\usepackage[dvipsnames,table]{xcolor}
\usepackage{slashed}
\usepackage{hyperref} 
\usepackage{xspace}
\usepackage[tight]{subfigure}
\usepackage{amsmath}
\usepackage{amsfonts}
\usepackage{mathrsfs}
\usepackage{comment}
\usepackage{afterpage}
\usepackage{verbatim}
\usepackage{booktabs}
\usepackage{array}
\usepackage[title,titletoc]{appendix}
\usepackage{tabularx}

\newcolumntype{C}[1]{>{\centering\arraybackslash}p{#1}}

\newcommand{\loqcd}{\rm LO_{\rm QCD}\xspace}
\newcommand{\loew}{\rm LO_{\rm EW}\xspace}
\newcommand{\nloone}{{\rm NLO}_1\xspace}
\newcommand{\nlotwo}{{\rm NLO}_2\xspace}
\newcommand{\nlothree}{{\rm NLO}_3\xspace}

\def\refeq#1{\mbox{(\ref{#1})}}
\def\reffi#1{\mbox{Fig.~\ref{#1}}}
\def\reffis#1{\mbox{Figs.~\ref{#1}}}
\def\refta#1{\mbox{Table~\ref{#1}}}

\def\refse#1{\mbox{Section~\ref{#1}}}
\def\refses#1{\mbox{Sections~\ref{#1}}}

\def\citere#1{\mbox{Ref.~\cite{#1}}}
\def\citeres#1{\mbox{Refs.~\cite{#1}}}

\newcommand{\newc}{\newcommand}
\newc{\beq}{\begin{equation}}
\newc{\eeq}{\end{equation}}
\newc{\bit}{\begin{itemize}}
\newc{\eit}{\end{itemize}}
\newc{\ben}{\begin{enumerate}}
\newc{\een}{\end{enumerate}}
\newc{\bce}{\begin{center}}
\newc{\ece}{\end{center}}
\newc{\bfi}{\begin{figure}}
\newc{\efi}{\end{figure}}

\newcommand{\rT}{{\mathrm{T}}}

\newcommand{\ie}{\emph{i.e.}\ }

\newcommand{\GeV}{\ensuremath{\,\text{GeV}}\xspace}
\newcommand{\TeV}{\ensuremath{\,\text{TeV}}\xspace}
\newcommand{\fb}{{\ensuremath\unskip\,\text{fb}}\xspace}

\newcommand{\PH}{\ensuremath{\text{H}}\xspace}

\newcommand{\Pp}{\ensuremath{\text{p}}}
\newcommand{\Pe}{\ensuremath{\text{e}}\xspace}
\newcommand{\Pb}{\ensuremath{\text{b}}\xspace}

\newcommand{\Pt}{\ensuremath{\text{t}}\xspace}
\newcommand{\Pu}{\ensuremath{\text{u}}\xspace}
\newcommand{\Pd}{\ensuremath{\text{d}}\xspace}
\newcommand{\Ps}{\ensuremath{\text{s}}\xspace}
\newcommand{\Pc}{\ensuremath{\text{c}}\xspace}
\newcommand{\Pg}{\ensuremath{\text{g}}}

\newcommand{\PW}{\ensuremath{\text{W}}\xspace}
\newcommand{\PZ}{\ensuremath{\text{Z}}\xspace}
\newcommand{\Pl}{\ell}
                                    
\newcommand{\Mt}{\ensuremath{m_\Pt}\xspace}
\newcommand{\MH}{\ensuremath{M_\PH}\xspace}

\newcommand{\Gt}{\ensuremath{\Gamma_\Pt}\xspace}
\newcommand{\GH}{\ensuremath{\Gamma_\PH}\xspace}

\newcommand{\recola}{{\sc Recola}\xspace}

\newcommand{\mocanlo}{{\sc MoCaNLO}\xspace}
\newcommand{\collier}{{\sc Collier}\xspace}

\newcolumntype{.}{D{.}{.}{-1}}
\newcolumntype{d}[1]{D{.}{.}{#1}}
\colorlet{tableoverheadcolor}{gray!37.5}
\colorlet{tableheadcolor}{gray!25}
\colorlet{tablerowcolor}{gray!12.5}

\def\draftdate{\relax}
\def\mda{\relax}
\def\mua{\relax}
\def\mla{\relax}
\def\draft{
\def\thtystars{******************************}
\def\sixtystars{\thtystars\thtystars}
\typeout{}
\typeout{\sixtystars**}
\typeout{* Draft mode!
         For final version remove \protect\draft\space in source file *}
\typeout{\sixtystars**}
\typeout{}
\def\draftdate{\today}
\def\mua{\marginpar[\boldmath\hfil$\uparrow$]%
                   {\boldmath$\uparrow$\hfil}\color{black}%
                    \typeout{marginpar: $\uparrow$}\ignorespaces}
\def\mda{\color{red}\marginpar[\boldmath\hfil$\downarrow$]%
                   {\boldmath$\downarrow$\hfil}%
                    \typeout{marginpar: $\downarrow$}\ignorespaces}
\def\mla{\marginpar[\boldmath\hfil$\rightarrow$]%
                   {\boldmath$\leftarrow $\hfil}%
                    \typeout{marginpar: $\leftrightarrow$}\ignorespaces}
\def\Mua{\marginpar[\boldmath\hfil$\Uparrow$]%
                   {\boldmath$\Uparrow$\hfil}\color{black}%
                    \typeout{marginpar: $\uparrow$}\ignorespaces}
\def\Mda{\color{red}\marginpar[\boldmath\hfil$\Downarrow$]%
                   {\boldmath$\Downarrow$\hfil}%
                    \typeout{marginpar: $\downarrow$}\ignorespaces}
\def\Mla{\marginpar[\boldmath\hfil\textcolor{red}{$\Rightarrow$}]%
                   {\boldmath\textcolor{red}{$\Leftarrow $}\hfil}%
                    \typeout{marginpar: $\leftrightarrow$}\ignorespaces}
\overfullrule 5pt
\oddsidemargin 15mm
\marginparwidth 29mm
}

\newcommand{\mc}{\mathcal}
\let\nnb\notag

\let\Mw\MW

\let\Mz\MZ

\let\Mwo\MWOS
\let\Gwo\GWOS
\let\Mzo\MZOS
\let\Gzo\GZOS

\let\as\alphas
\newcommand{\pt}[1]{p_{\rT,{#1}}}

\begin{document}
\title{
  Combined NLO EW and QCD corrections to off-shell $\Pt\overline{\Pt}\PW$ production at the LHC
}

\author{Ansgar Denner\and Giovanni Pelliccioli}
\institute{University of W\"urzburg,
  Institut f\"ur Theoretische Physik und Astrophysik,
  Emil-Hilb-Weg 22, 97074 W\"urzburg (Germany)}
\abstract{
  The high luminosity that will be accumulated at the LHC will enable precise
  differential measurements of the hadronic production of a top--antitop-quark
  pair in association with a $\PW$~boson. Therefore, an accurate description of
  this process is needed for realistic final states.
  In this work we combine for the first time the NLO QCD and electroweak
  corrections to the full off-shell $\Pt\overline{\Pt}\PW^+$ production at
  the LHC in the three-charged-lepton channel, including all spin~correlations,
  non-resonant effects, and interferences. 
  To this end, we have computed the NLO electroweak radiative corrections
  to the leading QCD order as well as the NLO QCD corrections to
  both the QCD and the electroweak leading orders. 
}
\authorrunning{\emph{A. Denner, G. Pelliccioli}}
\titlerunning{\emph{Combined NLO EW and QCD corrections to off-shell $\Pt\overline{\Pt}\PW$ production at the LHC}}
\maketitle

\section{Introduction}\label{intro}
The hadronic production of top-antitop pairs in association with a $\PW$~boson
is an interesting process to investigate at the Large Hadron Collider (LHC),
as it represents an important probe of the Standard Model (SM) as well as
a window to new physics.

This process is one of the heaviest signatures measurable at the LHC. It gives
access to the top-quark coupling to weak bosons and to possible deviations
from its SM value \cite{Dror:2015nkp,Buckley:2015lku,Bylund:2016phk}.
Due to the absence of a neutral initial state at a lower perturbative order than
next-to-next-to-leading order (NNLO) in QCD, it is also expected to improve
substantially the sensitivity to the $\Pt\overline{\Pt}$ charge asymmetry
\cite{Maltoni:2014zpa}.
Polarization observables and asymmetries in $\Pt\overline{\Pt}\PW^\pm$ production
are capable of enhancing the sensitivity to  beyond-the-SM (BSM)  interactions featuring
a chiral structure different from the one of the SM \cite{Maltoni:2014zpa,Bevilacqua:2020srb}.
The hadro-production of $\Pt\overline{\Pt}\PW^\pm$ is in general well suited
to directly search for BSM physics, in particular
supersymmetry \cite{Barnett:1993ea,Guchait:1994zk},
supergravity \cite{Baer:1995va}, technicolour \cite{Chivukula:1994mn},
vector-like quarks \cite{Aguilar-Saavedra:2013qpa}, Majorana neutrinos
\cite{Almeida:1997em} and modified Higgs sectors
\cite{Maalampi:2002vx,Perelstein:2005ka,Contino:2008hi}.
Beyond its own importance in LHC searches, the $\Pt\overline{\Pt}\PW$
production is a relevant background to $\Pt\overline{\Pt}{\rm H}$ production
\cite{Maltoni:2015ena}.

The ATLAS and CMS collaborations have measured and investigated
$\Pt \overline{\Pt}\PW^\pm $ production at Run~1 \cite{Aad:2015eua,Khachatryan:2015sha}
and Run~2 \cite{Aaboud:2016xve,Sirunyan:2017uzs,Aaboud:2019njj,CMS:2019too}  of the LHC.
This signature has been included as a background in the
recent experimental analyses for $\Pt\overline{\Pt}{\rm H}$ production
\cite{Aaboud:2018urx,Sirunyan:2018hoz,ATLAS-CONF-2019-045,CMS-PAS-HIG-17-004}.

The most recent experimental results based on Run~2
show a tension between data and theory predictions in the
$\Pt\bar{\Pt}\PW$ modelling both in direct measurements
\cite{Sirunyan:2017uzs,Aaboud:2019njj} and in the context of the search for
$\Pt\bar{\Pt}$ associated production with a Higgs boson
\cite{ATLAS-CONF-2019-045,CMS-PAS-HIG-17-004}. While the theoretical
community has invested a noticeable effort to address this tension, 
so far no explanation emerged that is capable to fill the gap
between the SM predictions and the data.

An improved modelling of the $\Pt\bar{\Pt}\PW^\pm$ process is required to allow
for the comparison of SM predictions with future LHC data,
particularly those that will be accumulated during the high-luminosity
run. The increased statistics will enable not only more precise
measurements of $\Pt\bar{\Pt}\PW^\pm$ cross-sections, but also measurements
of differential distributions and in different decay channels. This
target can only be achieved if the theoretical description of
realistic final states embedding the $\Pt\bar{\Pt}\PW$ resonance
structure is available.

Many theoretical predictions for $\Pt\overline{\Pt}\PW^\pm$
hadro-pro\-duc\-tion are available in the literature.  The first
next-to-leading order (NLO) QCD calculation for TeV-scale colliders
was performed in a spin-correlated narrow-width approximation for the
semi-leptonic decay channel \cite{Campbell:2012dh}. The matching of
NLO QCD predictions to a parton~shower was first tackled for the same
decay channel in \citere{Garzelli:2012bn}.  A number of calculations
for $\Pt\overline{\Pt}\PW^\pm$ inclusive production (on-shell
top/antitop quarks and $\PW$~boson) have been carried out, targeting charge
asymmetries \cite{Maltoni:2014zpa}, the impact of
$\Pt\overline{\Pt}\PW^\pm$ on the associated production of
$\Pt\overline{\Pt}$ pairs with a Higgs boson \cite{Maltoni:2015ena} at
NLO QCD, and the effects of subleading NLO QCD and electroweak (EW)
corrections \cite{Frixione:2015zaa,Frederix:2017wme,Frederix:2018nkq}.
Soft-gluon resummation up to next-to-next-to-leading logarithmic
accuracy
\cite{Li:2014ula,Broggio:2016zgg,Kulesza:2018tqz,Broggio:2019ewu,Kulesza:2020nfh}
and multi-jet merging \cite{vonBuddenbrock:2020ter} have also been
investigated. NLO QCD corrections to the EW leading order (LO) have
been computed in the narrow-width approximation (NWA), accounting for
complete spin correlations and including parton-shower effects
\cite{Frederix:2020jzp,1843174}.  Very recently a comparison of different
fixed-order Monte Carlo generators matched to parton~showers has been
performed for inclusive $\Pt\overline{\Pt}\PW^\pm$, with a focus on
the two-charged lepton signature (in the NWA) \cite{1843174}.
%
\begin{figure*} 
  \centering
  \includegraphics[scale=0.30]{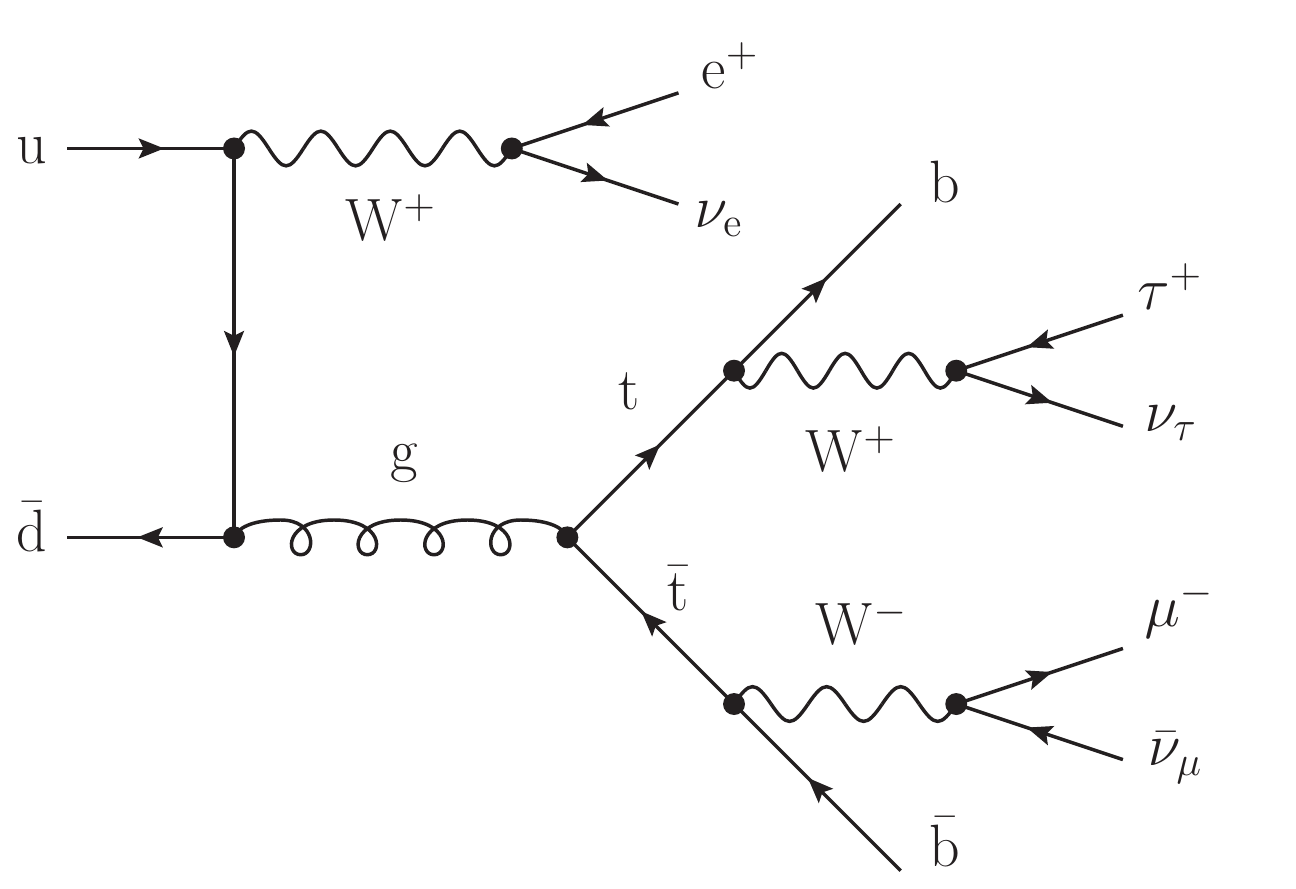}
  \includegraphics[scale=0.30]{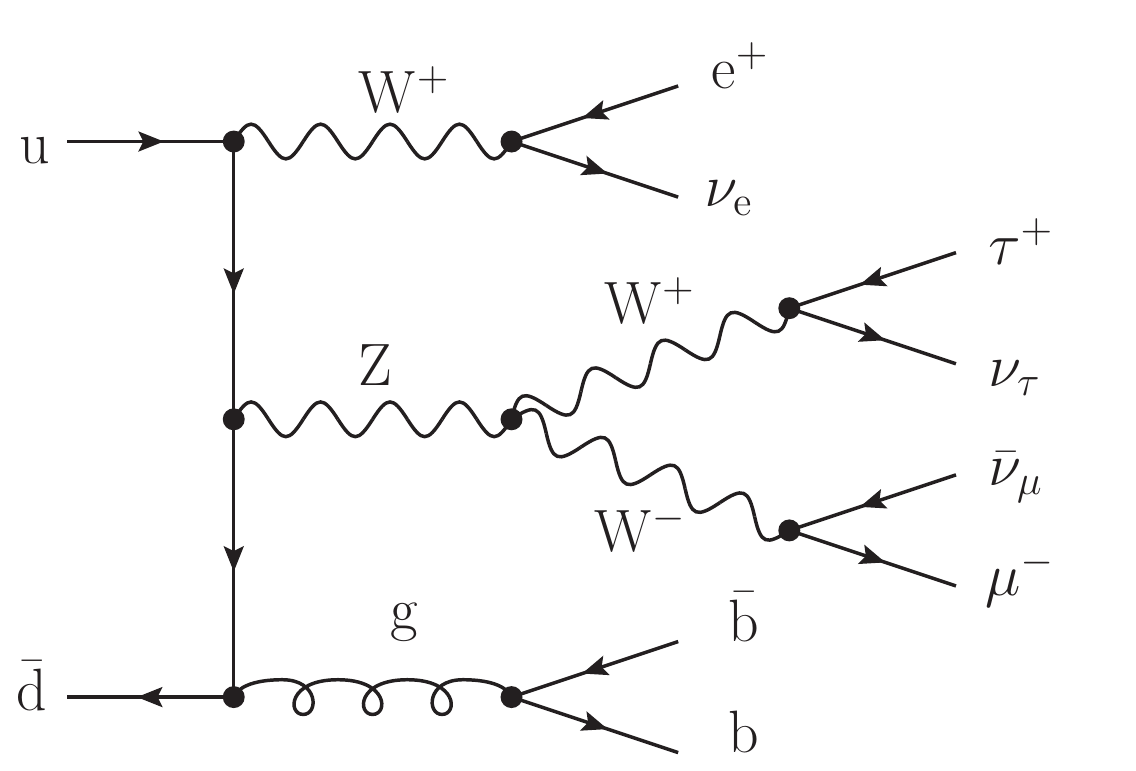}
  \includegraphics[scale=0.30]{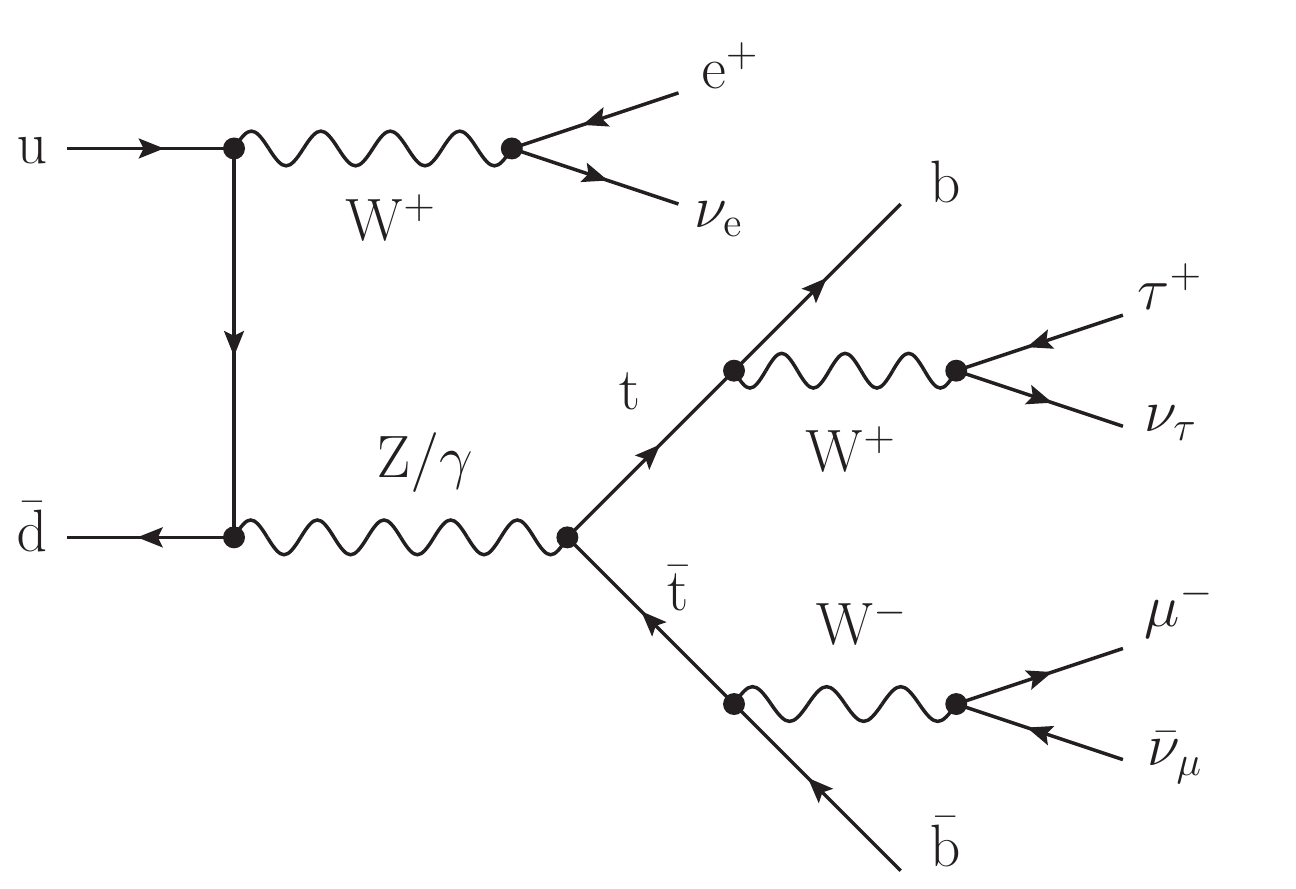}
  \includegraphics[scale=0.30]{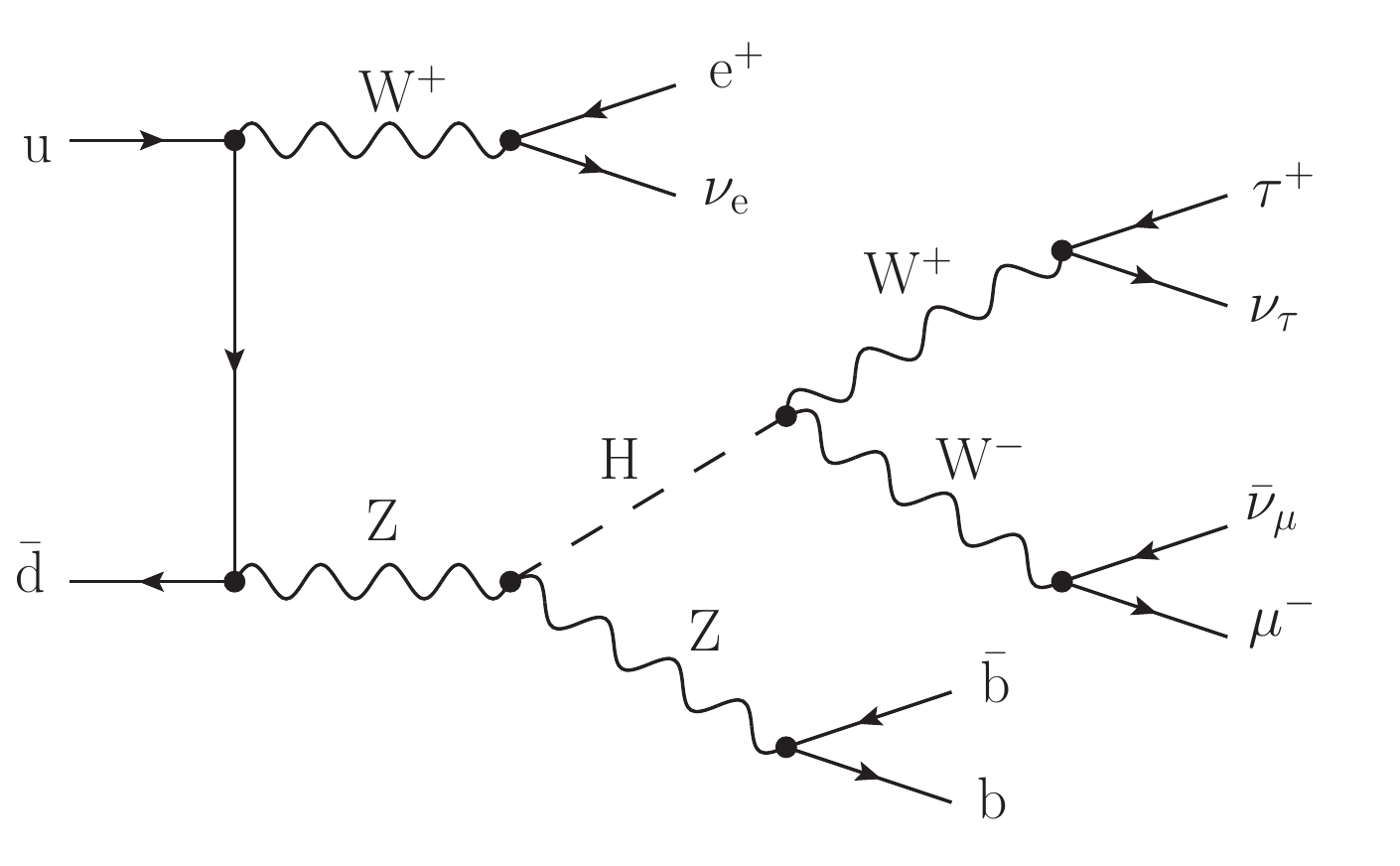}
  \caption{
    Sample diagrams contributing to $\loqcd$ (left) and to $\loew$ (right) cross-sections
    for off-shell $\Pt\overline{\Pt}\PW^+$ production in the three-charged-lepton channel.
  }\label{fig:lo}
\end{figure*}
%
\begin{figure*} 
  \centering
  \includegraphics[scale=0.6]{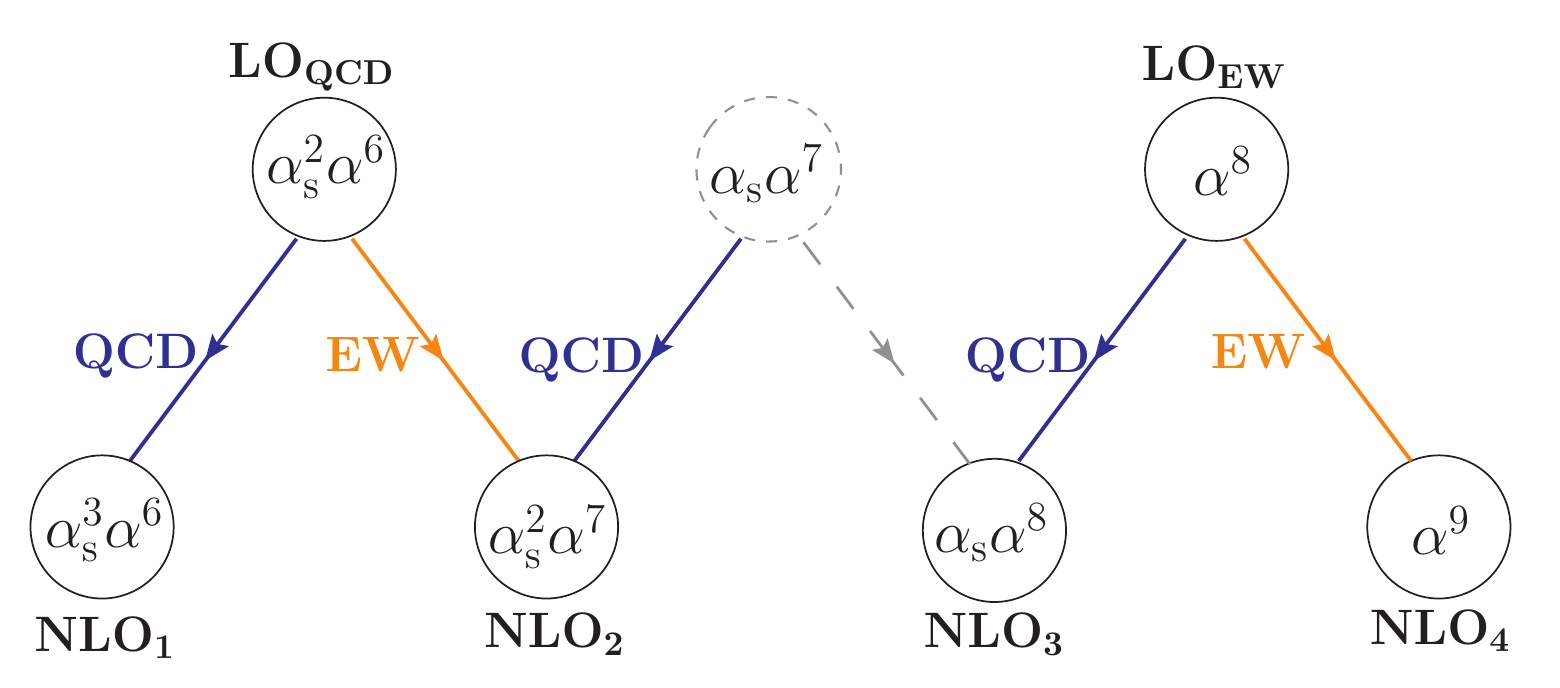}  
  \caption{
    Contributing perturbative orders at LO and NLO
    for $\Pt\overline{\Pt}\PW$ hadro-production in the three-charged-lepton channel.
  }\label{fig:orders}
\end{figure*}
%
The first studies that aim at the off-shell modelling of $\Pt\bar{\Pt}\PW$ production
concern the full NLO QCD corrections in the three-charged-lepton channel
\cite{Bevilacqua:2020pzy,Denner:2020hgg,Bevilacqua:2020srb}.

The calculation of subleading NLO corrections has been performed
for inclusive production \cite{Frixione:2015zaa,Frederix:2017wme,Frederix:2018nkq,Frederix:2020jzp,1843174},
but is still missing for realistic final states.
Our present work targets the complete fixed-order description
of the off-shell $\Pt\bar{\Pt}\PW$ production, combining the
NLO QCD and EW corrections that have a sizeable impact
at the LHC@13TeV.

This paper is organized as follows. In \refse{sec:descr} we describe the
process under investigation, providing details on various NLO corrections
that are presented. In \refses{sec:input} and \ref{sec:select} we provide
the SM input parameters and the selection cuts used for
numerical simulations, respectively. The integrated results at LO and NLO are presented
in \refse{subsec:integrated}, while in \refse{subsec:distrib} a number of
differential distributions are described, focusing
on the relative impact of various NLO corrections to the LO predictions.
In \refse{concl} we draw our conclusions.

\section{Details of the calculation}\label{details}

\subsection{Description of the process}\label{sec:descr}
We consider the process
\beq
\Pp\Pp\rightarrow \Pe^+\nu_{\Pe}\tau^+\nu_{\tau}\mu^-\bar{\nu}_\mu\,\Pb\,\bar{\Pb}\, + X\,,
\eeq
which receives contributions only from quark-induced partonic channels at LO.
Gluon--quark and photon--quark channels open up at NLO, while the pure gluonic
channel enters only at NNLO in QCD.

Although we consider the final state with three char\-ged leptons with different
flavours, the corresponding results for the case of identical positively-charged
leptons can be estimated by multiplying our results by a factor $1/2$, up to
interference contributions, which are expected to be small.

In this work we focus on the production of $\Pt\bar{\Pt}$ pairs in association
with a $\PW^+$ boson, but the calculation of the charge-conjugate process
($\Pt\bar{\Pt}\PW^-$) can be performed with the same techniques and no additional
conceptual issues.

At LO, the largest contribution is given by the QCD-mediated process of order
$\mc O(\as^2\alpha^6)$ (labelled $\loqcd$), which always embeds a gluon
$s$-channel propagator if no quark-family mixing is assumed (diagonal quark-mixing matrix
with unit entries).
The tree-level EW contribution of order $\mc O(\alpha^8)$ (labelled $\loew$),
despite being characterized by many more diagram topologies, is expected to give a
cross-section that is roughly 1\% of the $\loqcd$ one owing to the
ratio of EW and strong coupling constants.
The interference contribution, formally of order $\mc O(\as\alpha^7)$, is identically
zero due to colour algebra.
In \reffi{fig:lo} we show sample diagrams for the QCD-mediated and purely-EW process.
Note that the diagrams with a resonant top--antitop-quark pair are
present also in EW tree-level contributions.

At NLO, the $\Pt\overline{\Pt}\PW$ process receives contributions from four
different perturbative orders, as depicted in \reffi{fig:orders}.

The corrections that have the largest impact on the NLO cross-section
are of order $\mc O(\as^3\alpha^6)$, which are pure QCD corrections to
$\loqcd$. Following the notation of \citeres{Frederix:2017wme,Frederix:2018nkq}, we
label this perturbative order as $\nloone$. These corrections have recently
been computed
for the full off-shell process \cite{Bevilacqua:2020pzy,Denner:2020hgg}.
This perturbative order shows a
typical NLO QCD behaviour in the scale dependence, and the NLO relative corrections
to $\loqcd$ are at the $10$--$20\%$ level, depending on the choice of the
renormalization and factorization scale \cite{Denner:2020hgg}.

The $\nlotwo$ corrections are known only for on-shell top--antitop quarks and for
an on-shell $\PW$~boson \cite{Frixione:2015zaa,Frederix:2017wme,Frederix:2018nkq}.
They are expected to give a negative contribution of about 4.5\% of the inclusive
LO cross-section.

In the off-shell calculation, as well as in the on-shell one,
the $\nlotwo$ order receives contributions not only from the EW corrections to $\loqcd$,
but also from the QCD corrections to the LO interference, although at Born-level
the $\mc O(\as\alpha^7)$ contribution vanishes.

\begin{figure*} 
  \centering
  \includegraphics[scale=0.33]{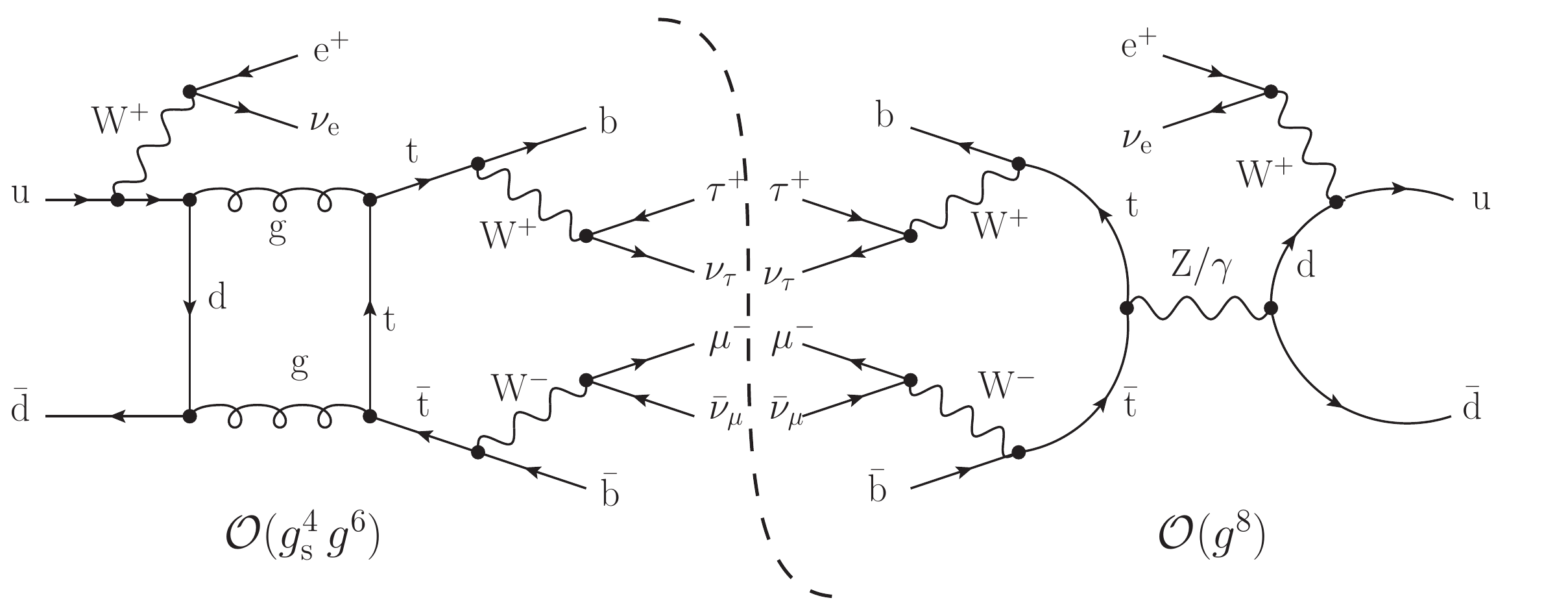}
  \includegraphics[scale=0.33]{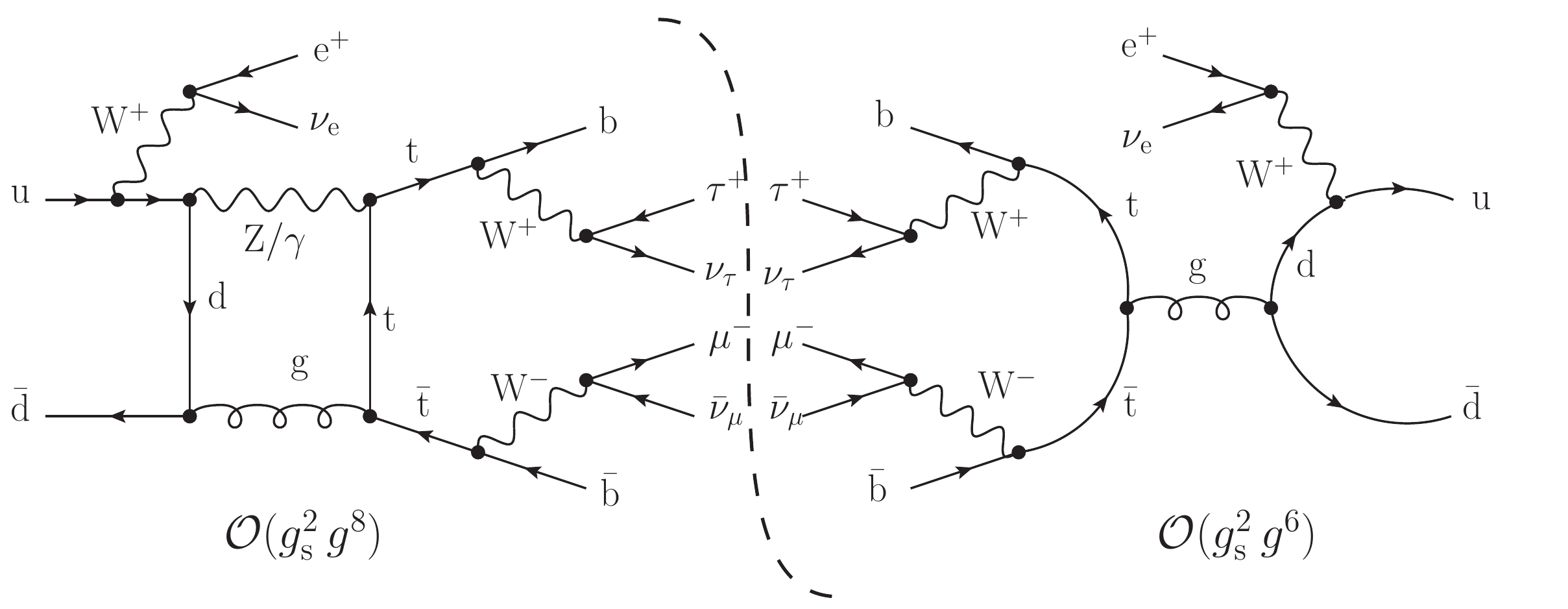}
  \caption{
    Sample contributions to the virtual corrections at order
    $\mc O(\as^2\alpha^7)$ for off-shell $\Pt\overline{\Pt}\PW$ production
    in the three-charged-lepton channel: QCD corrections to the
    LO interference (left) and a contribution that cannot be uniquely
    attributed to either the QCD corrections to the LO interference or the EW
    corrections to the $\loqcd$ (right). }\label{fig:nlovirtclasses}
\end{figure*}
%
\begin{figure*} 
  \centering
  \includegraphics[scale=0.33]{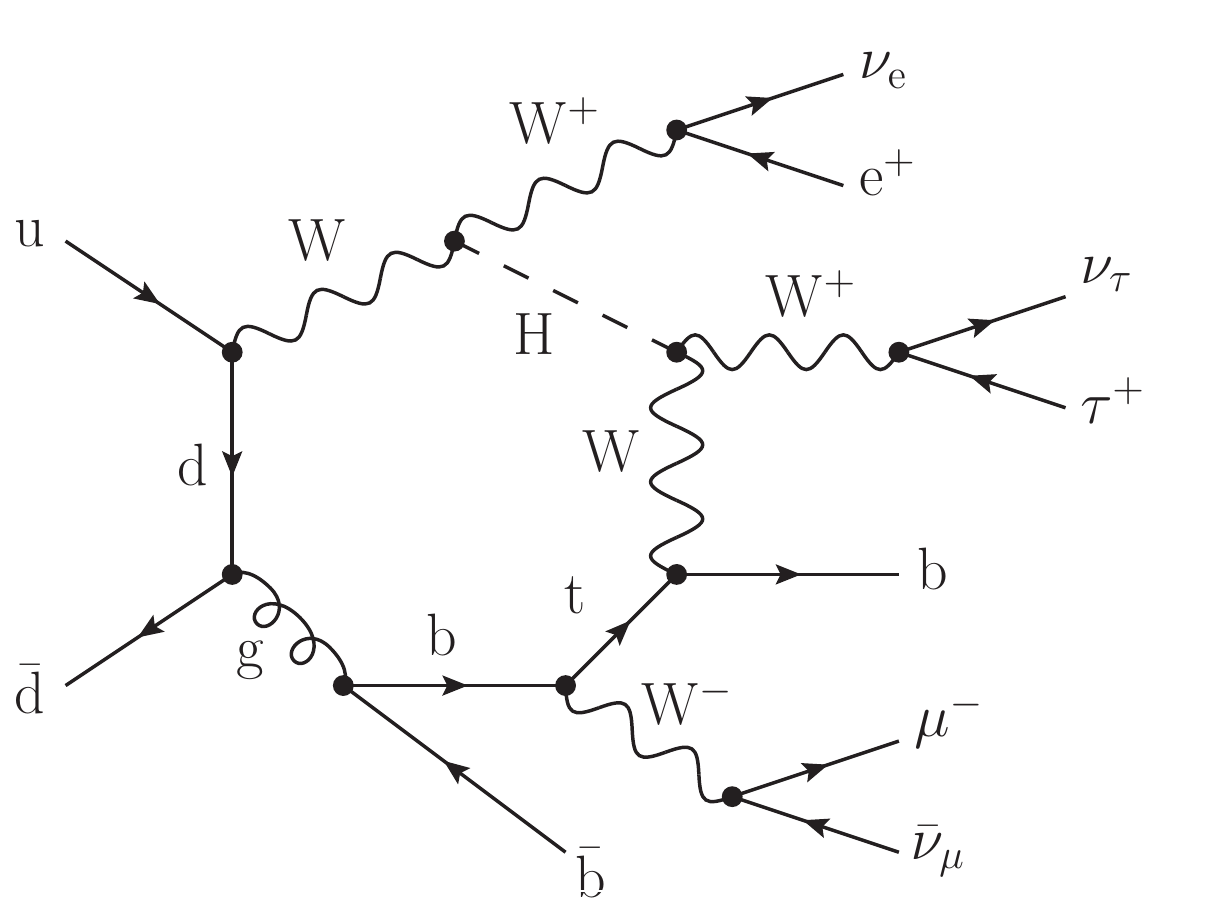}
  \includegraphics[scale=0.33]{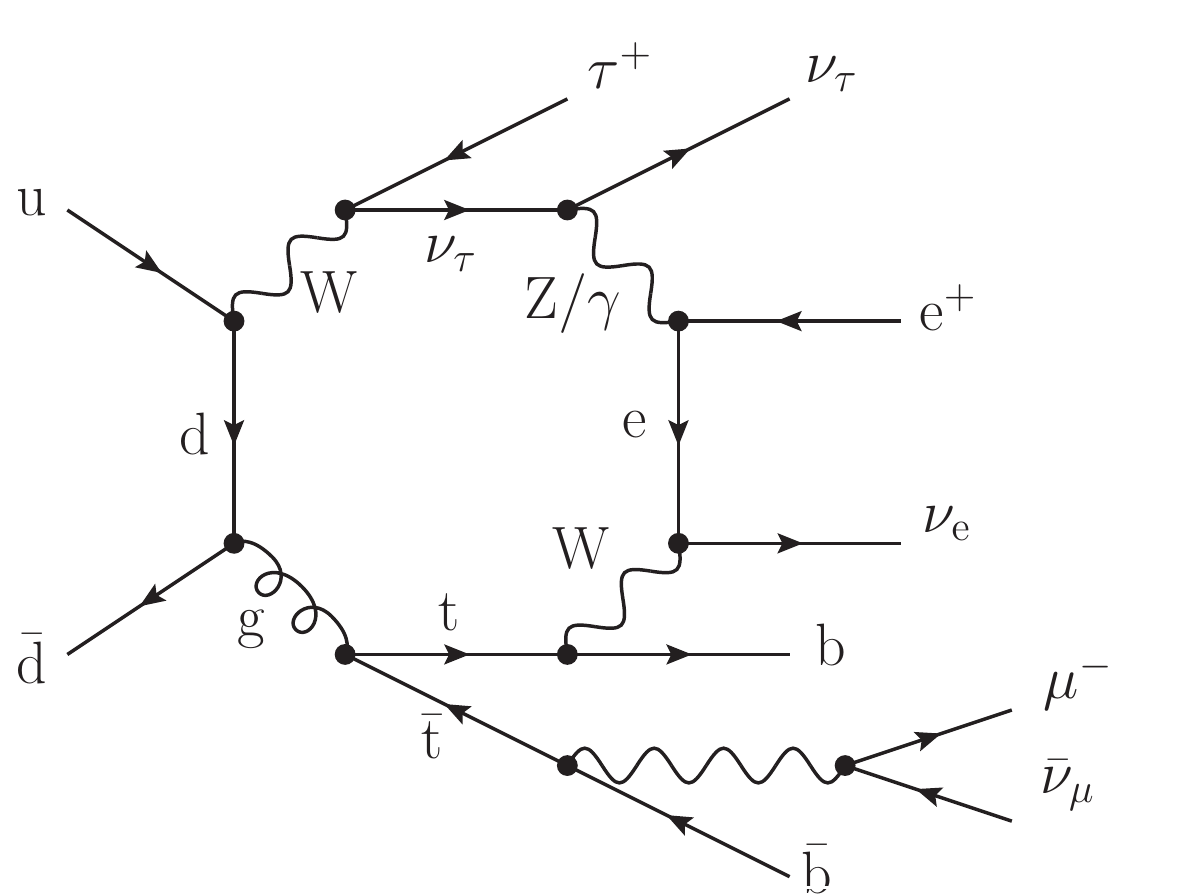}
  \includegraphics[scale=0.34]{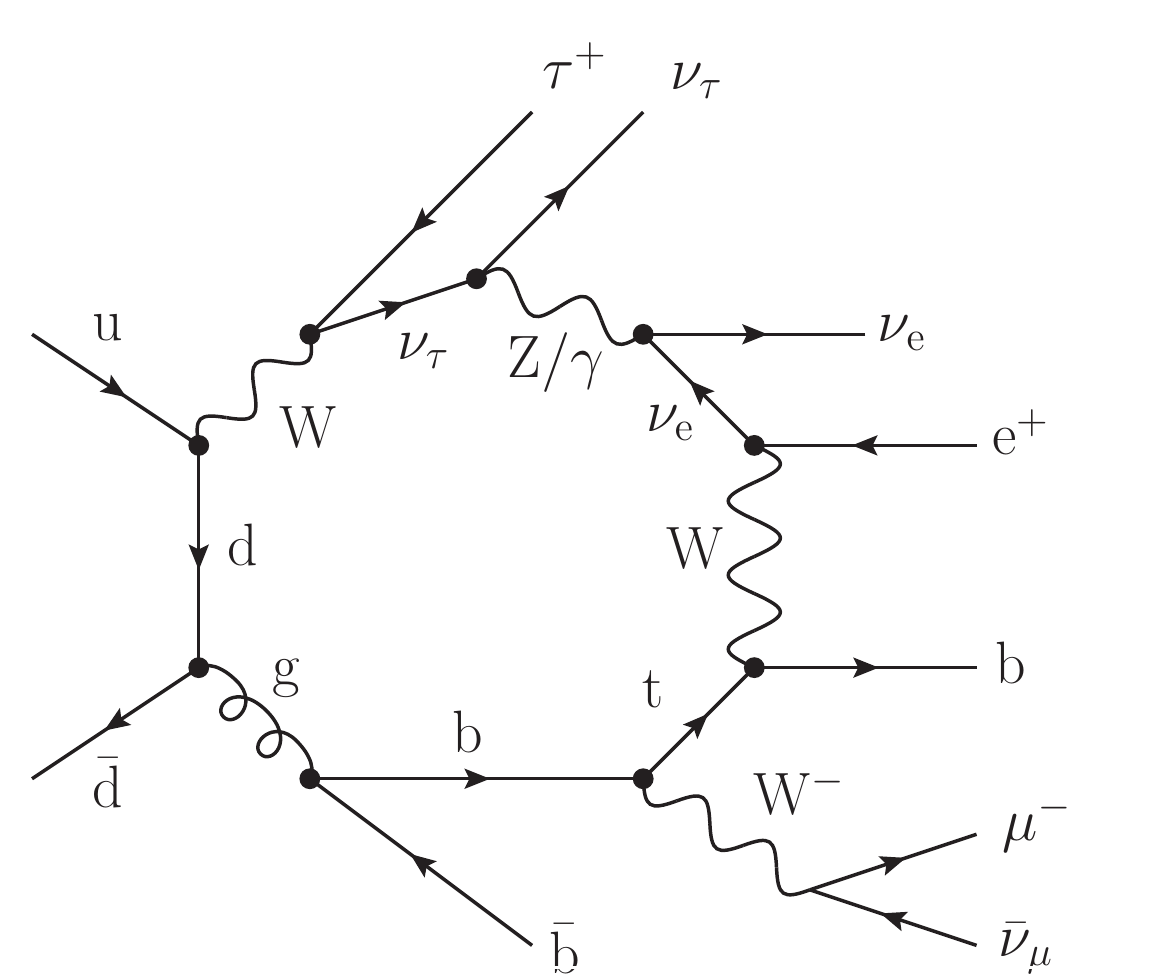}
  \includegraphics[scale=0.34]{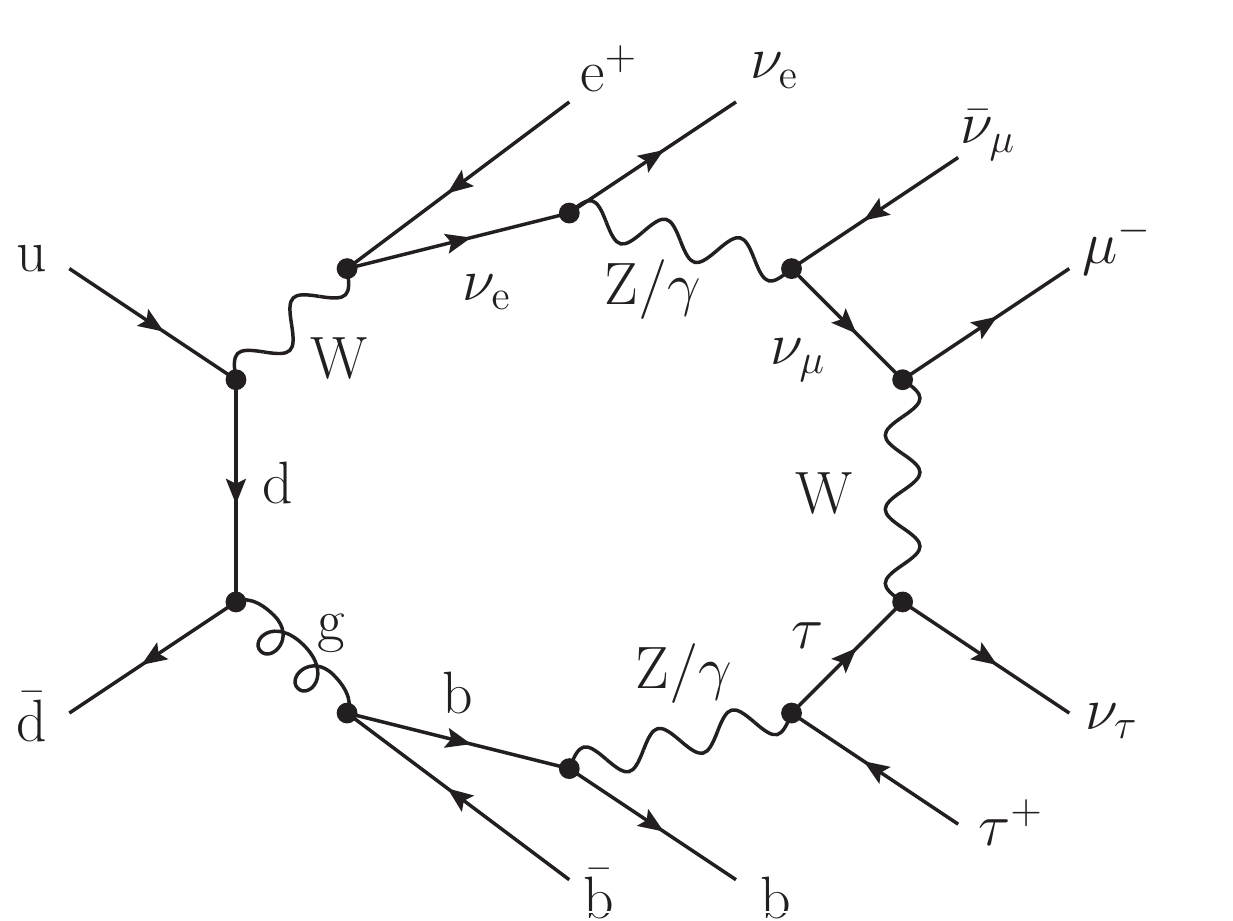}
  \caption{
    One-loop diagrams of order $\mc O(g_{\rm s}^2g^8)$ contributing to the EW virtual corrections ($\nlotwo$) to
   off-shell $\Pt\overline{\Pt}\PW$ production in the three-charged-lepton channel. From left to right: sample diagrams involving 7-, 8-, 9- and 10-point functions.}\label{fig:nlovirt}
\end{figure*}
%
Sample contributions to the virtual corrections at order $\mc
O(\as^2\alpha^7)$ are shown in \reffi{fig:nlovirtclasses}. The diagram on
the left involves one-loop amplitudes of order $\mc O(g_{\rm s}^4g^6)$
interfered with tree-level EW amplitudes of order $\mc O(g^8)$ and is
obviously a QCD correction to the LO interference. The diagram on the right
involves one-loop amplitudes of order $\mc O(g_{\rm s}^2g^8)$ interfered
with tree-level QCD amplitudes of order $\mc O(g_{\rm s}^2g^6)$ and could
be na\"ively classified as an EW correction to $\loqcd$.  However, it can
also be regarded as contributing to the QCD corrections to the LO
interference. In fact, the IR singularities of this contribution are
partially cancelled by the real photonic corrections to $\loqcd$ and
partially by the real gluonic corrections to the LO interference.  Diagrams
with weakly-interacting particles in the loops are more demanding from the
computational point of view, as the corresponding one-loop amplitudes include up to
10-point functions, while the latter ones feature at most
7-point functions. A selection of one-loop diagrams
which contribute at this perturbative order are shown in \reffi{fig:nlovirt}.

\begin{figure*}
  \centering
  \includegraphics[scale=0.33]{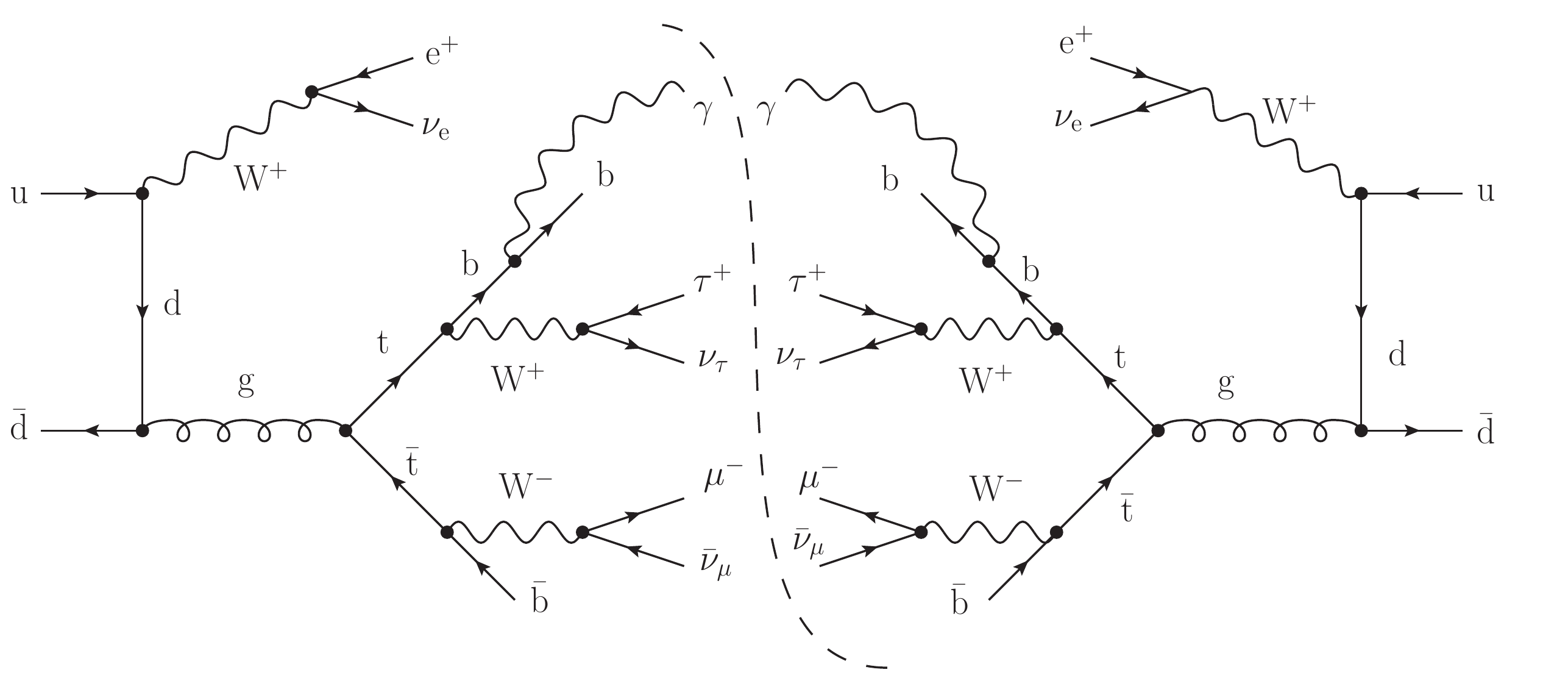}%
  \includegraphics[scale=0.33]{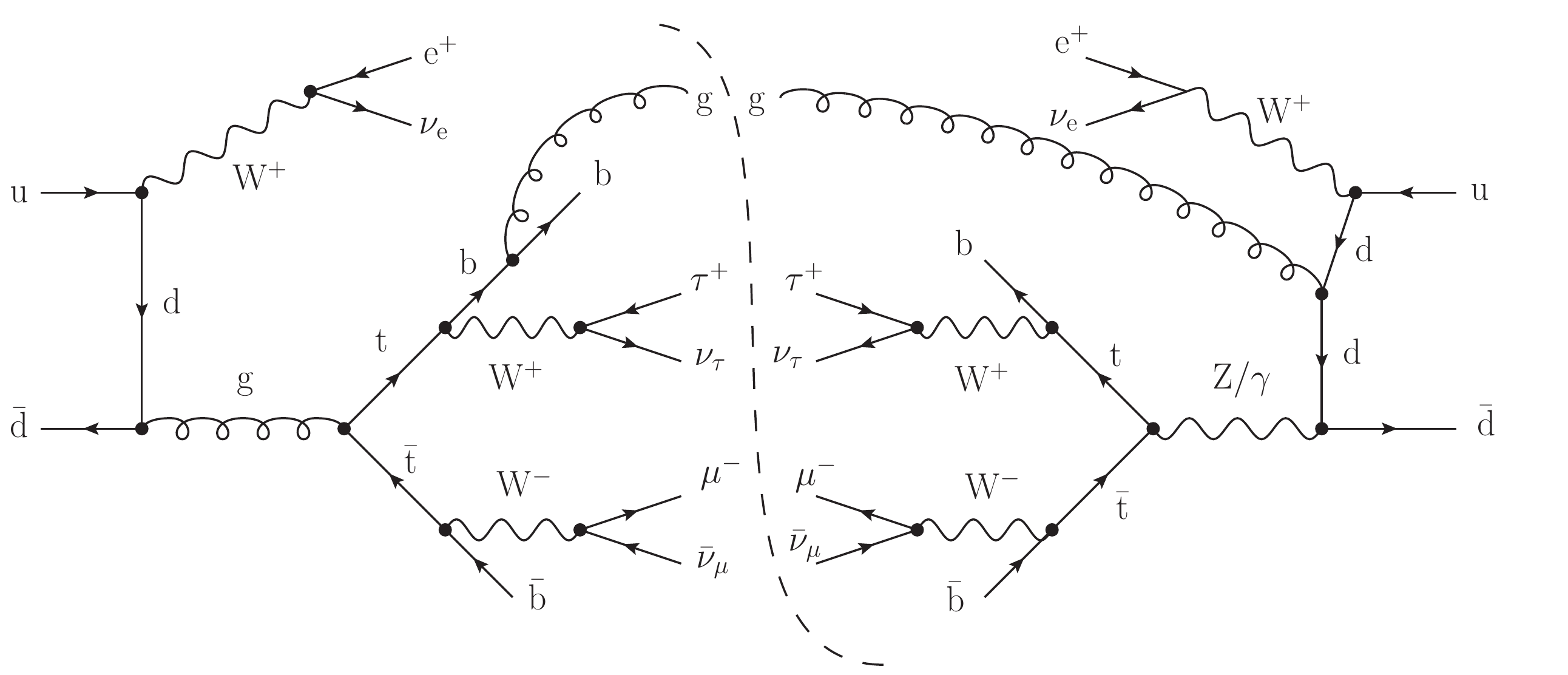}%
  \caption{
    Sample contribution to the real  corrections at order $\mc O(\as^2\alpha^7)$ for
    $\Pt\overline{\Pt}\PW$ production in the three-charged-lepton channel:
    photonic corrections to $\loqcd$(left) and gluonic corrections to the LO interference (right).
  }\label{fig:nlo2realinterf}
\end{figure*}

The real-radiation contributions to $\nlotwo$ corrections are
computationally demanding due to the large multiplicity of electrically
charged final-state particles.  In contrast to the virtual ones, the real $\nlotwo$
corrections can be uniquely classified into  two types: the NLO EW corrections to the $\loqcd$
process, which involve a real photon (see \reffi{fig:nlo2realinterf} left
for an example), and the NLO QCD corrections to the LO interference, which
involve a real gluon (see \reffi{fig:nlo2realinterf} right for an example).  In the
first class of contributions, the photon can be either in the final or in
the initial state. The processes with a photon in the final state are
characterized by many singular regions, as the photon can become soft or
collinear to any of the seven charged external particles.  This results in
a large number of subtraction counterterms that are required to ensure a stable
calculation of the NLO cross-section.  The real processes with a
photon in the initial state possess a smaller number of singular
phase-space regions and are suppressed due to the small luminosity of
photons in the proton.  For on-shell production, the contribution of the
photon-induced channels to the leading-order cross-section is at the
sub-percent level \cite{Frixione:2015zaa}.  The QCD corrections to the LO
interference are non-vanishing only if the radiated gluon is emitted by an
initial-state light quark and absorbed by a final-state b quark
or top quark (or the other way around). A sample contribution is shown in
\reffi{fig:nlo2realinterf} right.
These corrections, 
although necessary to account for all $\mc O(\as^2\alpha^7)$ contributions,
turn out to be very small, as detailed in \refse{results}.

To sum up, the full set of real partonic channels that contribute
to the $\nlotwo$ corrections is
\beq
\left.
\begin{array}{ll}
  u\,\bar{d}  \rightarrow  \Pe^+\nu_{\Pe}\, \mu^-\bar{\nu}_\mu\, \tau^+\nu_{\tau}\, \Pb \, \bar{\Pb} \,  \gamma \,\\
  \gamma \, u \rightarrow      \Pe^+\nu_{\Pe}\, \mu^-\bar{\nu}_\mu\, \tau^+\nu_{\tau}\,\Pb \, \bar{\Pb} \,  d \,\\
  \gamma \, \bar{d} \rightarrow \Pe^+\nu_{\Pe}\, \mu^-\bar{\nu}_\mu\, \tau^+\nu_{\tau}\,\Pb \, \bar{\Pb} \,  \bar{u} \, 
\end{array}
\right\}
    {\textrm{EW corr. to $\loqcd$}}\nnb
\eeq
and
\begin{equation}
 \left.
\begin{array}{ll}
  u\,\bar{d}  \rightarrow  \Pe^+\nu_{\Pe}\, \mu^-\bar{\nu}_\mu\, \tau^+\nu_{\tau}\, \Pb \, \bar{\Pb} \,  \Pg \,\\
  \Pg \, u  \rightarrow     \Pe^+\nu_{\Pe}\, \mu^-\bar{\nu}_\mu\, \tau^+\nu_{\tau}\,\Pb \, \bar{\Pb} \,  d \,\\
  \Pg \, \bar{d} \rightarrow \Pe^+\nu_{\Pe}\, \mu^-\bar{\nu}_\mu\, \tau^+\nu_{\tau}\,\Pb \, \bar{\Pb} \,  \bar{u} \, 
\end{array}
\right\}
  {\textrm{QCD corr. to LO int.,}}\nnb
\end{equation}
where $u$ and $d$ stand for up-type and down-type quarks, respectively
(of the first and second generation). 

\begin{figure*} 
  \centering
  \includegraphics[scale=0.34]{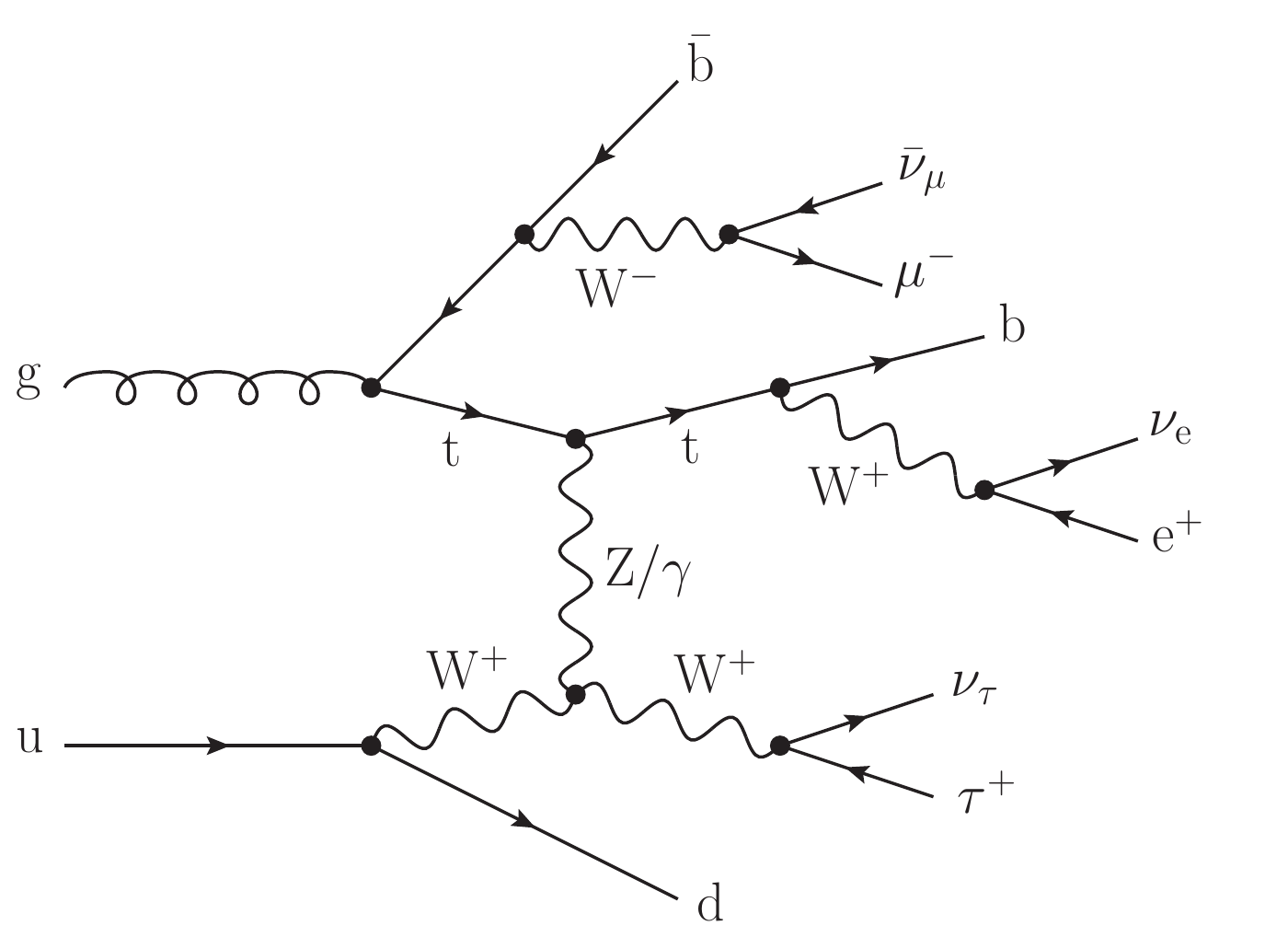}
  \includegraphics[scale=0.34]{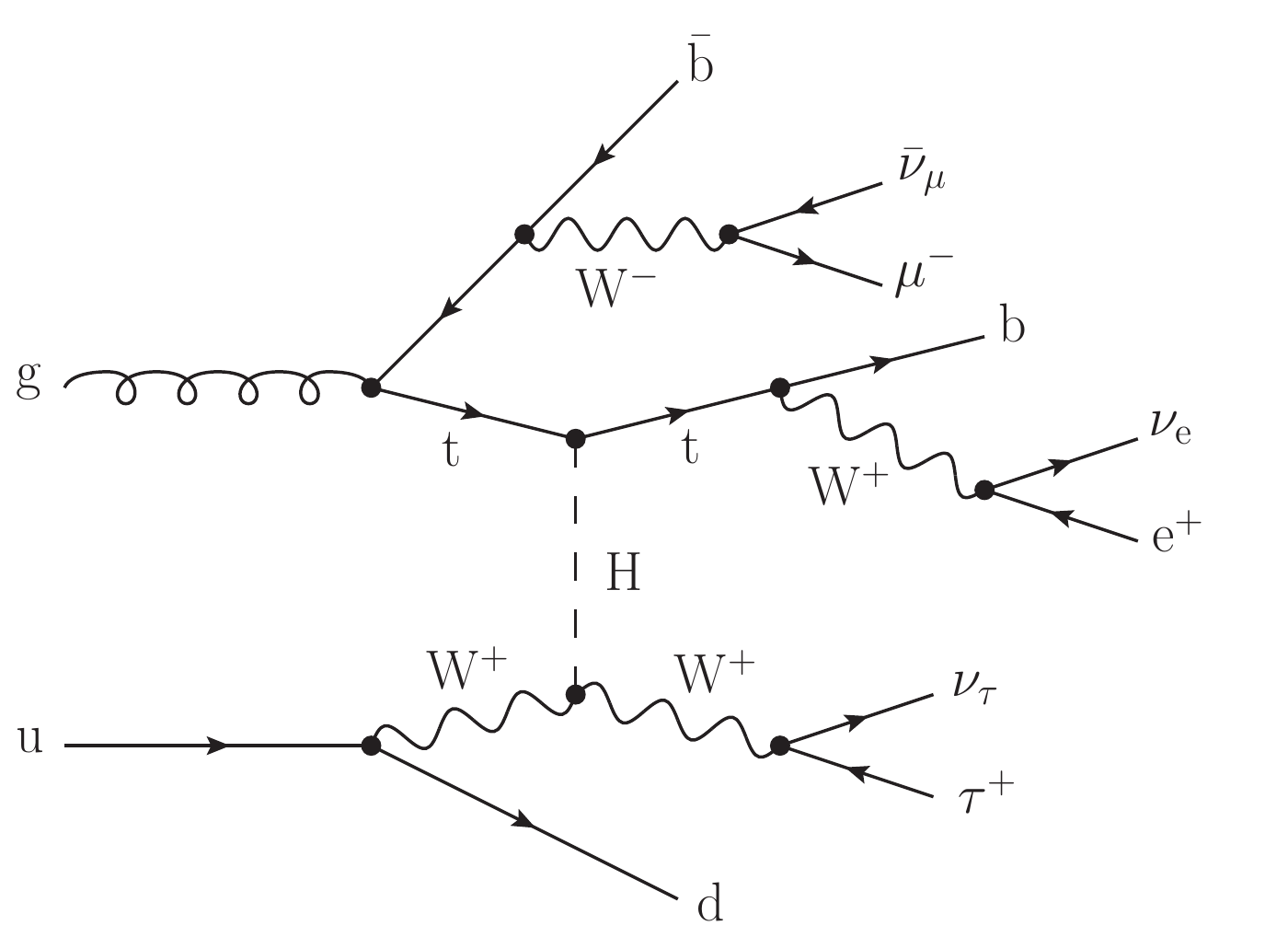}
  \includegraphics[scale=0.34]{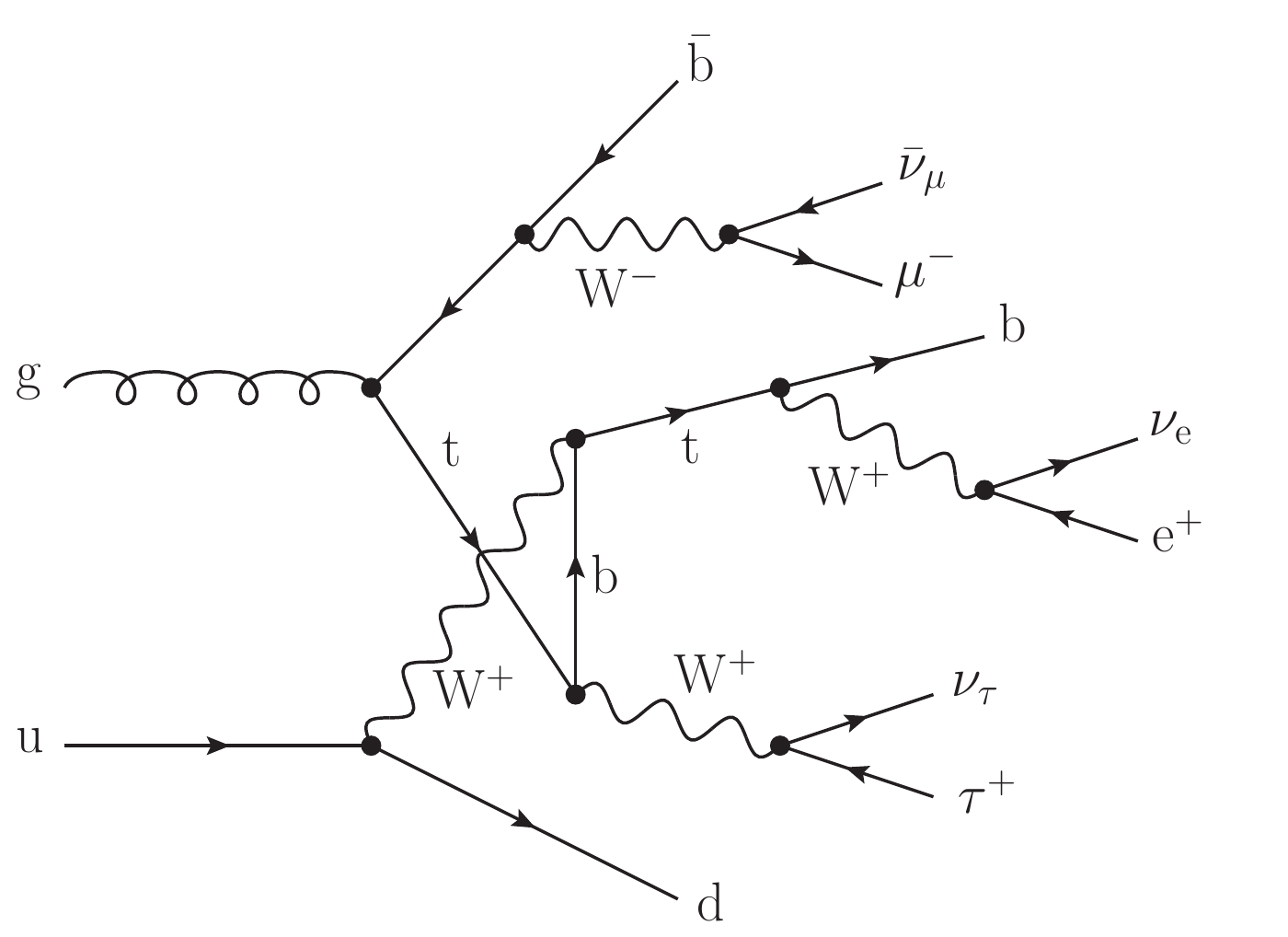}
  \caption{Sample diagrams for the partonic channel $\Pg\Pu\rightarrow \Pb\,\bar{\Pb}\,\Pe^+\nu_{\Pe}\tau^+\nu_{\tau}\mu^-\bar{\nu}_\mu \Pd$ that
    contain $\Pt\PW^+$ scattering as a subprocess and
    contribute to $\nlothree$ corrections to $\Pt\overline{\Pt}\PW$
    production in the three-charged-lepton
    channel.}\label{fig:scatttw} 
\end{figure*}

The vanishing LO interference implies that the corresponding EW corrections vanish as well,
since additional EW propagators (virtual contributions) and radiated photons
(real contributions) do not modify the LO colour structure.
Therefore, the only NLO corrections that contribute
at order $\mc O(\as\alpha^8)$ are genuine QCD corrections to the $\loew$ cross-section.
This order is labelled as $\nlothree$ in \reffi{fig:orders}.
By simply counting the powers of $\as$ the $\nlothree$ corrections are expected to give a
smaller contribution than the $\nlotwo$ ones.
However, at the inclusive level \cite{Frederix:2017wme} and in the narrow-width
approximation \cite{Frederix:2020jzp,1843174}, they are noticeably larger than the $\nlotwo$ ones.
This results from the fact that this perturbative order is dominated by hard
real radiation diagrams in the gluon-quark partonic channel that embed the $\Pt\PW$ scattering
process \cite{Dror:2015nkp}. Sample diagrams are shown in \reffi{fig:scatttw}.
Thanks to the genuine QCD nature of the $\nlothree$ corrections,
it is possible to match them to a QCD parton~shower with
no subtleties due to EW corrections, as it has been done
in \citeres{Frederix:2020jzp,1843174}.

The last NLO perturbative order, $\mc O(\alpha^9)$, is furnished by the
EW corrections to the $\loew$ process. It has been shown at the inclusive
level that such contributions are at the sub-permille level \cite{Frederix:2018nkq},
as expected by na\"ive power counting. Even with a substantially larger 
data set than the one of Run~2 (\ie 3000 $\rm fb^{-1}$ at the
high-luminosity LHC)
these EW effects are out of reach in a realistic fiducial region.
Therefore, we are not providing results for this order.  

In the following we focus on the first three NLO perturbative orders.
Tree-level and one-loop SM amplitudes are computed with the \recola
matrix-element provider \cite{Actis:2012qn,Actis:2016mpe}. For the tensor reduction and evaluation
of loop integrals we use the \collier library \cite{Denner:2016kdg}.
The multi-channel Monte Carlo integration is performed with \mocanlo,
a generator that has already been used to compute the NLO QCD corrections to
$\Pt\overline{\Pt}\PW$ \cite{Denner:2020hgg} and the NLO EW corrections
to several LHC processes involving top quarks \cite{Denner:2016jyo,Denner:2016wet}.
The subtraction of infrared and collinear singularities is carried out
using the dipole formalism of \citeres{Catani:1996vz,Dittmaier:1999mb,Catani:2002hc}
both for QCD and for EW corrections.
The initial-state collinear singularities are absorbed in the parton
distribution functions (PDFs) in the $\overline{\rm MS}$ factorization scheme.

\subsection{Input parameters}\label{sec:input}
We consider proton--proton collisions at a centre-of-mass energy of 13 TeV.
We neglect flavour mixing in the quark sector and use a unit quark-mixing matrix.
The three charged leptons that we consider in the final state are massless
and characterized by three different flavours.
The on-shell masses and widths of weak bosons
are set to the following values \cite{Tanabashi:2018oca},
\begin{align}
\Mwo ={}& 80.379 \GeV,\qquad\,\,\, \Gwo = 2.085\GeV\,, \nnb\\
\Mzo ={}& 91.1876 \GeV,\qquad \Gzo \,= 2.4952\GeV\,,
\end{align}
and then translated into their pole values \cite{Bardin:1988xt}
that enter the Monte Carlo simulations.
The Higgs-boson mass and width are fixed, following \citere{Tanabashi:2018oca},
to
\begin{align}
\MH  ={}& 125 \GeV, \qquad\,\,\, \GH\, \,\,= 0.00407\,\GeV\,.
\end{align}
We have computed the LO top-quark width according to \citere{Jezabek:1988iv},
using the pole values for the $\PW$-boson mass and width.
The NLO top-quark width is obtained applying the NLO QCD and EW correction
factors of \citere{Basso:2015gca} to the LO width. The numerical values read
\begin{align}
\Mt ={}& 173.0\GeV\,,& \nnb\\
\Gt^{\rm LO}   ={}& 1.4437\GeV\,, \quad
\Gt^{\rm NLO}  = 1.3636\GeV\,.&
\end{align}
The top-quark width is kept fixed when performing var\-iations of the
factorization and renormalization scales.
The EW coupling is treated in the $G_\mu$~scheme \cite{Denner:2000bj}, 
\beq
\alpha = \frac{\sqrt{2}}{\pi}\,G_\mu\Mw^2\left[1-\frac{\Mw^2}{\Mz^2}\right]\,,
\eeq
where { $G_\mu = 1.16638\cdot10^{-5} \GeV^{-2}$} is the Fermi constant.
The masses of weak bosons and of the top quark, and therefore also
the EW mixing angle, are treated in the complex-mass scheme
\cite{Denner:1999gp,Denner:2000bj,Denner:2005fg,Denner:2006ic}.

Both for LO and NLO predictions, we employ \sloppy NNPDF\_3.1\_luxQED
PDFs \cite{Bertone:2017bme}.
Using these PDFs, the photon
contribution is properly accounted for in the evolution.
The evaluation of PDFs and the running of the strong coupling 
are obtained via the {\scshape LHAPDF6} interface
\cite{Buckley:2014ana}. The employed PDF set uses $\as(\Mz)=0.118$ and one
QCD loop in the calculation of the $\as$ evolution.

\subsection{Selection cuts}\label{sec:select}
Coloured partons with $|\eta|<5$ are clustered into jets by means of
the anti-$k_t$ algorithm \cite{Cacciari:2008gp} with resolution
radius $R=0.4$. The same algorithm but with $R=0.1$ is applied to
cluster photons into charged particles.

We choose cuts that mimic the fiducial selections applied by ATLAS in \citere{Aaboud:2019njj}
and that have already been used for the study of NLO QCD corrections to the same final
state \cite{Denner:2020hgg}. We select events with exactly
two $\Pb$~jets (assuming perfect b-tagging efficiency) that are required to satisfy
\beq
\pt{\Pb} > 25 \GeV\,, \quad |\eta_{\Pb} |<2.5\,.
\eeq
Furthermore, we ask for three charged leptons that fulfil standard
acceptance and isolation cuts,
\beq
\pt{\Pl} > 27 \GeV\,, \quad |\eta_\Pl |<2.5\,, \quad \Delta R_{\Pl \Pb}>0.4\,,
\eeq
where the $R$~distance is defined as the sum in quadrature of the
azimuthal and rapidity separations,
\beq\label{eq:Rdist}
\Delta R_{ij} = \sqrt{\Delta\phi_{ij}^{\,2}+\Delta y_{ij}^{\,2}}\,.
\eeq
We do not constrain the missing transverse momentum and do not apply any veto
to additional light jets. 

\section{Results}\label{results}
For the factorization and renormalization scale ($\mu_{\rm F} = \mu_{\rm R}$), we
consider two different dynamical choices that have proved to behave better
than a fixed scale \cite{Bevilacqua:2020pzy,Denner:2020hgg}.
The first one, introduced in \citere{Bevilacqua:2020pzy}, depends on the
transverse-momentum content of the final-state particles, regardless of
the top--antitop resonances,%
\footnote{We use the same notation for the scales as in \citere{Denner:2020hgg}.}
\beq\label{eq:htdef}
\mu_0^{\rm (c)} = \frac{H_{\rT}}{3} = \frac{\pt{\rm miss}+\sum_{i = {\Pb,\Pl}} \pt{i}}{3} \,.
\eeq
The second dynamical choice, already used in \citere{Denner:2020hgg},
is based on the transverse masses of the top and antitop quarks. Due to the
ambiguity in choosing the $\ell^+\nu_\ell$ pair that results from the top quark,
we pick the pair of leptons that when combined with the bottom quark forms
an invariant mass closest to the top-quark mass.
We consider two different central-scales based on this dynamical choice,
\beq
\mu_0^{\rm (d)}
=\sqrt{\!\sqrt{\Mt^2+\pt{\Pt}^{\,2}}\,\sqrt{\Mt^2+\pt{\overline{\Pt}}^{\,2}}}\,,
\qquad \mu_0^{\rm (e)} = \frac{\mu_0^{\rm (d)}}{2}\,. 
\eeq
The scale-dependence study performed in
\citere{Denner:2020hgg} shows that using $\mu_0^{\rm (e)}$ as a
central scale reduces the scale dependence based on the conventional
7-point scale variation and gives smaller QCD corrections than using
$\mu_0^{\rm (d)}$.  Therefore, the choice $\mu_0^{\rm (e)}$ is
preferable for the study of the impact of $\nlotwo$ and
$\nlothree$ corrections, which is the focus of this work.
  
The scale uncertainties shown in the following results are based on
7-point scale variations, \ie rescaling the central factorization
and renormalization scale by the factors
\begin{align}
\left(0.5,0.5\right),\left(1,0.5\right),\left(0.5,1\right),\left(1,1\right),\left(1,2\right),\left(2,1\right),\left(2,2\right)\,,\nnb
\end{align}
while keeping the NLO QCD top-quark width fixed.

In the following we present the results combining NLO corrections with the
so-called \emph{additive} approach,
\begin{align}
  \sigma_{\rm LO+NLO} ={}& \sigma_{\rm LO_{\rm QCD}}\,+\,\sigma_{\rm NLO_1}\,+\,\sigma_{\rm NLO_2} \nnb \\
 &+\sigma_{\rm LO_{\rm EW}}\,+\,\sigma_{\rm NLO_3}.
\end{align}
This approach is exact at the order of truncation of the perturbative expansion.
Furthermore, it represents a natural choice for our process,
as the combination involves NLO corrections to two different leading orders that do not interfere.

\subsection{Integrated cross-sections}\label{subsec:integrated}

In \refta{table:sigmainclNLO_combined}, we present the integrated cross-sections
in the fiducial region defined in \refse{sec:select}.
\begin{table*}
\begin{center}
\renewcommand{\arraystretch}{1.3}
\begin{tabular}{C{2.8cm}|C{2.cm}C{1.8cm}|C{2.cm}C{1.8cm}|C{2.cm}C{1.8cm}}%
  \hline
& \multicolumn{2}{c|}{$\mu_0^{\rm (c)}$ } & \multicolumn{2}{c|}{$\mu_0^{\rm (d)}$ } & \multicolumn{2}{c}{$\mu_0^{\rm (e)}$ } \\[0.5ex]
  \hline
  \hline
perturbative order  & $\sigma$ (fb)   &  ratio  & $\sigma$ (fb)   &  ratio  & $\sigma$ (fb)   &  ratio    \\[0.5ex] 
\hline
\hline
  $\rm LO_{\rm QCD}$ ($\as^2\alpha^6$) &   0.2218(1)$^{+25.3\%}_{-18.8\%}$  &  1  & 0.1948(1)$^{+23.9\%}_{-18.1\%}$  & 1 & 0.2414(1)$^{+26.2\%}_{-19.3\%}$  & 1  \\[0.5ex]          
  $\rm LO_{\rm EW}$ ($\alpha^8$)       &   0.002164(1)$^{+3.7\%}_{-3.6\%}$   &  0.010 & 0.002122(1)$^{+3.7\%}_{-3.6\%}$ & 0.011 & 0.002201(1)$^{+3.7\%}_{-3.6\%}$& 0.009  \\[0.5ex]
  \hline
  ${\rm NLO_1}$ ($\as^3\alpha^6$)   &   0.0147(6)    &  0.066  & 0.0349(6) & 0.179 & 0.0009(7)& 0.004  \\[0.5ex] 
  ${\rm NLO_2}$ ($\as^2\alpha^7$)&$\!\!$-0.0122(3)&$\!\!$-0.055 & $\!\!$-0.0106(3)& $\!\!$-0.054 & $\!\!$-0.0134(4) & $\!\!$-0.056  \\[0.5ex]
  ${\rm NLO_3}$ ($\as\alpha^8$)   &   0.0293(1)   &    0.131 & 0.0263(1)& 0.135 & 0.0320(1)&  0.133 \\[0.5ex]
  \hline
  $\rm LO_{\rm QCD}$+${\rm NLO_1}$ & 0.2365(6)$^{+2.9\%}_{-6.0\%}$ & 1.066& 0.2297(6)$^{+5.5\%}_{-7.3\%}$ & 1.179& 0.2423(7)$^{+3.5\%}_{-5.2\%}$ & 1.004\\[0.5ex]
  $\rm LO_{\rm QCD}$+${\rm NLO_2}$ & 0.2094(3)$^{+25.0\%}_{-18.7\%}$ &0.945 & 0.1840(3)$^{+23.8\%}_{-17.9\%}$& 0.946& 0.2277(4)$^{+25.9\%}_{-19.2\%}$ & 0.944\\[0.5ex]
  $\rm LO_{\rm EW}$+${\rm NLO_3}$ & 0.03142(4)$^{+22.2\%}_{-16.8\%}$ & 0.141 & 0.02843(6)$^{+20.5\%}_{-15.6\%}$ & 0.146& 0.03425(7)$^{+22.8\%}_{-17.0\%}$ &0.142\\[0.5ex]
  \hline
  \hline
  LO+NLO & 0.2554(7)$^{+4.0\%}_{-6.5\%}$ & 1.151 & 0.2473(7)$^{+6.3\%}_{-7.6\%}$ & 1.270 & 0.2628(9)$^{+4.3\%}_{-5.9\%}$ & 1.089\\[0.5ex]
  \hline
\end{tabular}
\end{center}
\caption{
  LO cross-sections and NLO corrections (in fb) in the fiducial setup, for three different dynamical-scale choices.
  Numerical errors (in parentheses) are shown. Ratios are relative to the $\rm LO_{\rm QCD}$ cross-section.
  The scale uncertainties from 7-point scale variations (in percentage) are listed for LO and NLO cross-sections.
  The result in the last row {is the sum} of all LO cross-sections and NLO corrections, namely
  $\rm LO_{\rm QCD}+ \rm LO_{\rm EW}+ {\rm NLO_1}+{\rm NLO_2}+{\rm NLO_3}$.
}\label{table:sigmainclNLO_combined}
\end{table*}

The leading corrections to the $\loqcd$ cross-section are expected to
come from the corresponding pure QCD radiative corrections ($\nloone$).
Their inclusion has been proved to decrease the theoretical uncertainty
due to scale variations and to stabilize the perturbative convergence
for this process. Nonetheless, their impact depends on the choice of the
scale, as already pointed out both in inclusive
\cite{Frixione:2015zaa,Frederix:2017wme} and off-shell \cite{Bevilacqua:2020pzy,Denner:2020hgg} computations.
In fact, with the resonance-blind dynamical choice $\mu_0^{\rm (c)}$,
the $\nloone$ corrections give a $6.6\%$ enhancement to the $\loqcd$ cross-section, with
the resonance-aware choices, $\mu_0^{\rm (d)}$ and $\mu_0^{\rm (e)}$, they give a $18\%$
and a $0.4\%$ correction,  respectively.
%
Note that, at variance with \citere{Denner:2020hgg} in this paper we compute
both LO and NLO predictions with the NLO top-quark width, which gives roughly
a $12\%$ enhancement to the LO cross-section.
%

The $\nlotwo$ corrections are negative and amount to about
$-5.5\%$ of the $\loqcd$ cross-section for all scale
choices. Such supposedly subleading corrections have a sizeable
impact on the fiducial NLO cross-section, and this is likely
due to large EW Sudakov logarithms enhancing the cross-section
in the high-energy regime \cite{Denner:2000jv}.
This is supported by the fact that the average partonic
centre-of-mass energy and $H_{\rm T}$ are quite high, $850\GeV$
and $520\GeV$, respectively. A crude estimate of the Sudakov
logarithms
gives a result
which is of the same order of magnitude as the 
full $\nlotwo$ corrections we have obtained for this process.

The impact of QCD corrections that can be uniquely attributed to the LO
interference is very small, both for real and for virtual corrections,
accounting respectively for 5\% and for less than 1\% of the total
$\nlotwo$ result.

At $\mc{O}(\as^2\alpha^7)$, we have also included the contribution from
photon-initiated partonic channels, which are positive and account for
about $0.1\%$ of the LO QCD cross-section. As already observed at the
inclusive level \cite{Frixione:2015zaa}, this contribution is very small
and its effect will be hardly visible even at the high-luminosity run of
LHC (they will yield about 1 event for $\sqrt{s}=13$ TeV and
$\mathcal{L}=3000\,{\rm fb^{-1}}$).

In inclusive calculations with on-shell top--antitop quarks,
the $\nlotwo$ corrections were found to give a $-4.5\%$ contribution to
the inclusive production cross-section \cite{Frixione:2015zaa,Frederix:2018nkq}.
In order to 
compare our results with those of \citere{Frederix:2018nkq},
we have performed a full off-shell calculation in a very inclusive
setup, and divided by the branching ratios for the decays of the 
top and antitop quarks and of the $\PW$~boson.
The setup is the same as the one of \citere{Frederix:2018nkq}, up to
a few unavoidable differences:
\begin{itemize}
\item we use finite top-quark and $\PW$ widths
  (same values as those of \refse{sec:input}) and we include Higgs-boson
  contributions;
\item we apply a minimum invariant-mass
  cut of $5\GeV$ to the $\rm b\bar{\rm b}$ system to protect from infrared singularities
  and we cluster photons into charged particles with isolation radius $R=0.1$;
\item we employ the same dynamical scale as in \citere{Frederix:2018nkq},
  but using the kinematics after photon recombination and choosing the
  top-quark candidate with the same invariant-mass prescription as for the
  calculation of the scales $\mu_0^{\rm (d)}$ and $\mu_0^{\rm (e)}$.
\end{itemize}
The obtained inclusive cross-sections,
\begin{align}
\sigma_{\rm LO_{\rm QCD}} ={}& 262^{+23.7\%}_{-18.1\%} \fb,\notag\\ 
\sigma_{\rm LO_{\rm QCD}+NLO_2}={}& 254^{+23.4\%}_{-17.8\%} \fb\,,
\end{align}
exhibit $\nlotwo$ corrections of $-3\%$ of $\loqcd$, which is
not far from the $-4.5\%$ of \citeres{Frixione:2015zaa,Frederix:2018nkq}.
The remaining discrepancy should be due to both the additional cuts
we have applied and the non-resonant effects that are included
in our full calculation, while being absent in on-shell calculations.
The comparison of results in the fiducial and the inclusive setup reveals
that the  $\nlotwo$ corrections are more sizeable for 
realistic final states and in the presence of reasonable fiducial cuts.

Coming back to our default fiducial setup (see \refse{sec:select}),
the $\nlothree$ perturbative order is dominated by the real radiation in the $\Pu \Pg$
partonic channel, owing to a PDF enhancement and the $\Pt\PW$
scattering embedded in this channel. These real corrections account
for 85\% of the $\nlothree$ corrections, and are one order of magnitude larger
than the corresponding leading order $\loew$.
The total $\nlothree$ corrections amount to $13\%$ of the $\loqcd$ cross-section,
almost independently of the scale choice.
This confirms the $12\%$ effect obtained in the case of on-shell top and
antitop quarks \cite{Frederix:2018nkq}.

It is worth stressing that the inclusion of $\nlotwo$ and $\nlothree$
corrections gives a noticeable effect to the $\Pt\bar{\Pt}\PW$ cross-section.
Therefore such corrections must definitely be accounted for in experimental
analyses. Furthermore, their relative contribution to the LO result
is rather independent of the scale choice, while the $\nloone$
corrections are much more scale dependent.


As a last comment of this section, we point out that the scale
uncertainty of the combined LO+NLO cross-section is driven
by the $\nloone$ corrections which reduce the $\loqcd$ uncertainty roughly
from $20\%$ to $5\%$. Due to their EW nature, the $\nlotwo$
corrections do not diminish the scale uncertainty of the corresponding
leading order ($\loqcd$). The $\nlothree$ corrections exceed the
corresponding  pure EW LO process and, thus,
imply a LO-like scale dependence for $\rm LO_{\rm EW}$+${\rm NLO_3}$.

So far, we have focused on the relative contributions of various
NLO corrections to the fiducial $\Pt\bar{\Pt}\PW$ cross-section. However,
the interplay among different corrections can be rather different for
more exclusive observables. Therefore, it is
essential to study differential distributions.

\begin{figure*} 
  \centering
    \subfigure[Transverse momentum of the positron.\label{fig:pte}]{\includegraphics[scale=0.4]{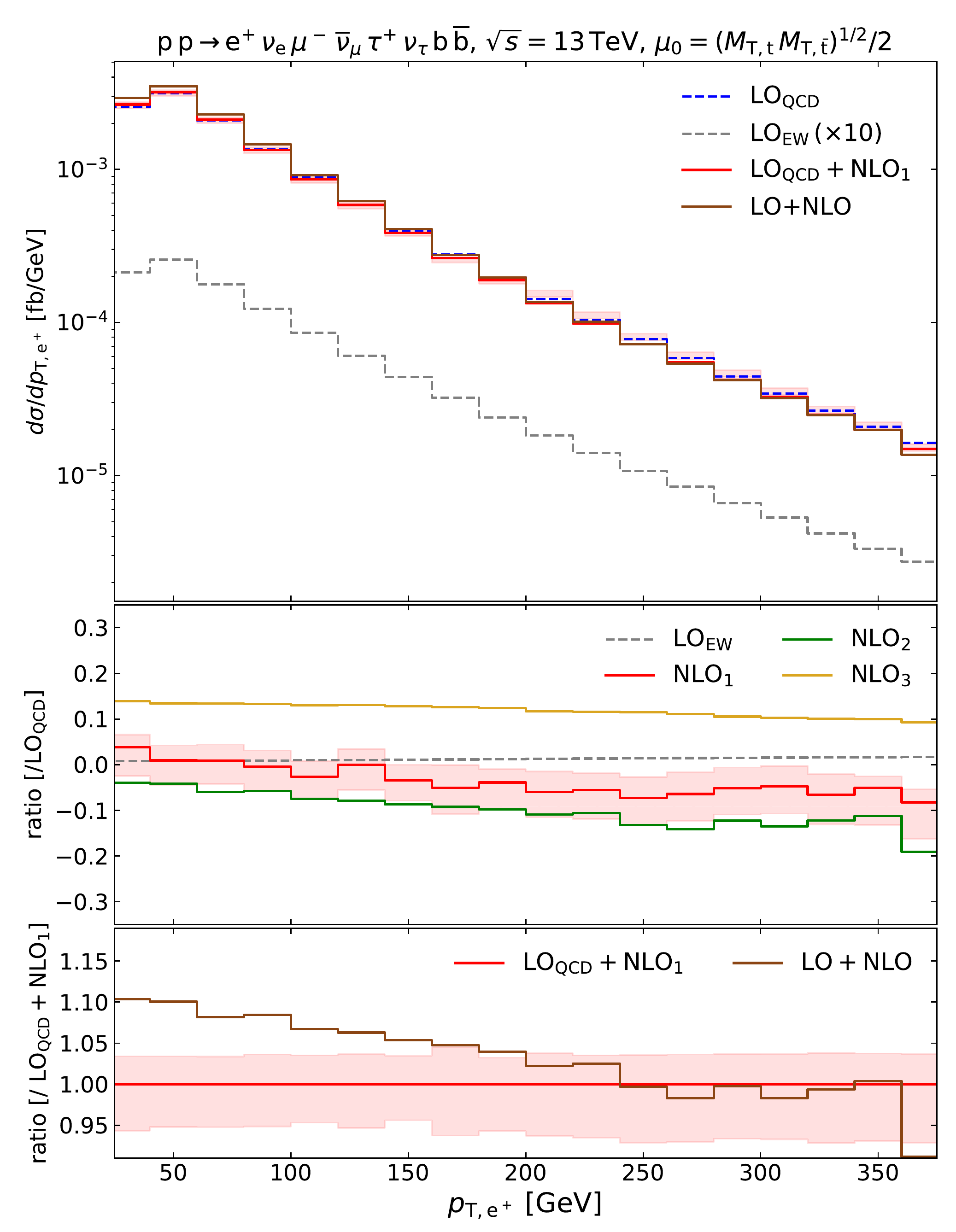} }
    \subfigure[Transverse momentum of the antitop quark.\label{fig:pttbar}]{\includegraphics[scale=0.4]{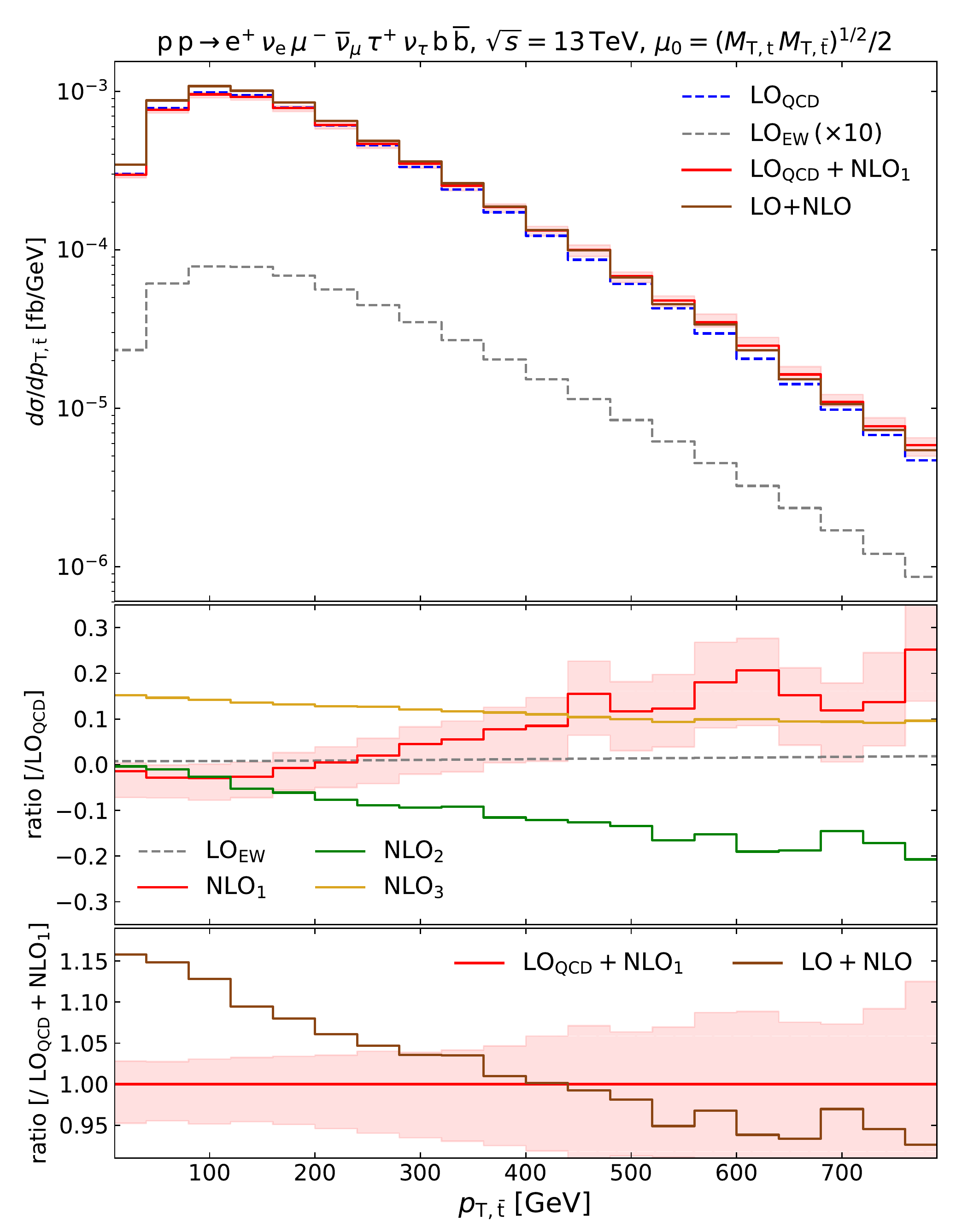} }
  \caption{Distributions in the transverse momentum of the positron (left)
    and of the antitop quark (right).
    Top panel: differential cross-sections (in fb)
    for $\rm LO_{\rm QCD}$, $\rm LO_{\rm EW}$  (scaled by a factor 10), $\rm LO_{\rm QCD}+NLO_1$ and
    for the complete NLO, which is the sum of all LO cross-sections
    and all NLO corrections. Middle panel: ratio of the $\rm LO_{\rm EW}$, $\rm NLO_1$,
    $\rm NLO_2$, and $\rm NLO_3$ corrections over the $\rm LO_{\rm QCD}$ cross-section.
    Bottom panel: ratio of the $\rm LO+NLO$ cross-section over the $\rm LO_{\rm QCD}+NLO_1$ one.
    Uncertainties from 7-point scale
    variations are shown in all panels for the $\rm LO_{\rm QCD}+NLO_1$ predictions.    
  }\label{fig:ptet}
\end{figure*}

\subsection{Distributions}\label{subsec:distrib}
In the following we present a number of relevant distributions focusing
on the impact of the various NLO corrections relative to the $\loqcd$ cross-sections.
Since in most of the LHC experimental analyses the theoretical predictions
are NLO QCD accurate, we also comment on the distortion of NLO QCD distribution shapes ($\loqcd+\nloone$ in
our notation) due to the inclusion of $\nlotwo$ and $\nlothree$ corrections.
We choose to present all differential distributions with the $\mu_0^{\rm(e)}$
scale.%
\footnote{
  In \citere{Denner:2020hgg} the scale
  for the differential distributions is $\mu_0^{\rm(d)}$, \ie
  exactly twice the default scale used here.
}
The shown scale uncertainties  are
based on the predictions for $\loqcd+\nloone$, normalized to
predictions at the central scale in the relative plots.
         
\begin{figure*} 
  \centering
  \subfigure[Transverse momentum of the $\Pt\bar{\Pt}$ system.\label{fig:pttt}]{\includegraphics[scale=0.4]{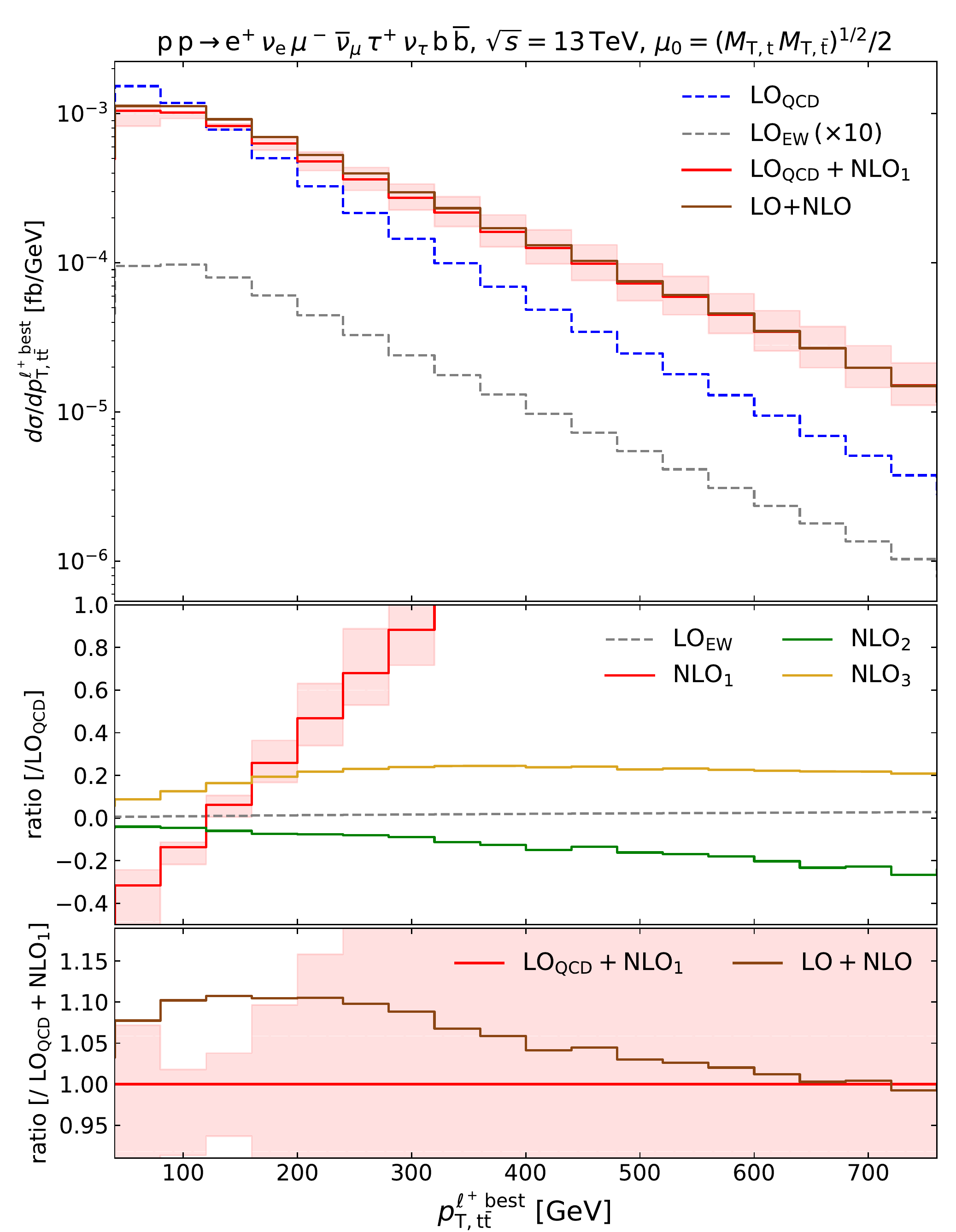}}
  \subfigure[Transverse momentum of the $\Pb\bar{\Pb}$ system.\label{fig:ptbb}]{\includegraphics[scale=0.4]{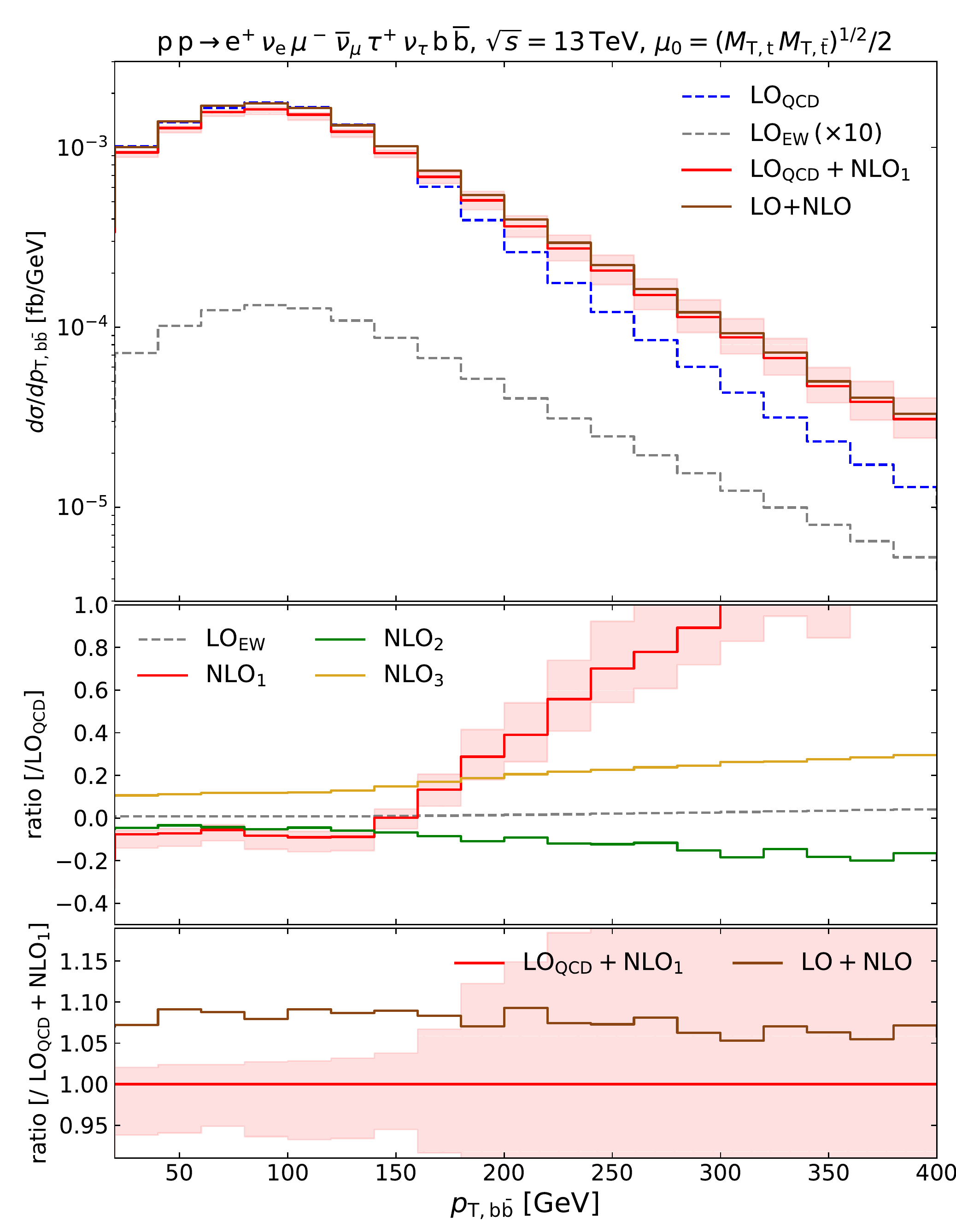}}
  \caption{Distributions in the transverse momentum of the reconstructed $\Pt\bar{\Pt}$ system (left)
    and of the $\Pb\bar{\Pb}$ system (right). Same structure as \reffi{fig:ptet}.}\label{fig:ptbbtt}
\end{figure*}

We start by presenting transverse-momentum distributions
in \reffis{fig:ptet}--\ref{fig:ptbbtt}.

In \reffi{fig:pte} we consider the distribution in the transverse momentum of the positron, which
is precisely measurable at the LHC.
Since we also include EW corrections ($\nlotwo$), the positron
is understood as dressed (a radiated photon
could be clustered into the positron).
The distribution peaks around $50\GeV$, where the relative impact of QCD and EW
corrections follows straight the integrated results.
Relative to the $\loqcd$, all three radiative corrections drop
in a monotonic manner. Nonetheless, the decrease of $\nlothree$ corrections
is very mild ($14\%$ below $50\GeV$, $10\%$ at $380\GeV$), while the
$\nloone$ and $\nlotwo$ corrections decrease steeper: the
former become negative around $80\GeV$ and give $-9\%$ at $380\GeV$,
the latter become lower than $-10\%$ already at moderate $\pt{\Pe^+}$ ($200\GeV$).
The behaviour of EW corrections in the tails of this distribution
is likely driven by the impact of Sudakov logarithms, which become large at high
$p_{\rm T}$.

The same behaviour of the $\nlotwo$ and $\nlothree$ corrections characterizes also the
distribution in the transverse momentum of the antitop quark, shown
in \reffi{fig:pttbar}.
The antitop momentum is computed as the sum of the momenta of the muon, its
corresponding antineutrino, and the antibottom quark. This is not observable
at the LHC, but its analysis is useful to compare the full off-shell calculation
with the on-shell ones. The negative growth of the $\nlotwo$ corrections behaves
very similarly in the inclusive calculations, as can be seen for example
in \citere{Frixione:2015zaa} (figure 5 therein). This confirms that for
sufficiently inclusive variables the NLO EW effects are dominated
by contributions with resonant top and antitop quarks.
The $\nloone$ corrections increase by roughly $25\%$ 
in the considered spectrum. 

In both transverse-momentum distributions of \reffi{fig:ptet}, the inclusion
of subleading NLO corrections ($\nlotwo,\nlothree$) gives a decreasing effect
towards large transverse momenta to the NLO QCD cross-section.
In fact, the ratio between the combined $\rm LO+NLO$ cross-section and the
$\loqcd+\nloone$ one ranges between $1.10$ and $1.15$ for small transverse momenta
and drops below $1$ already at moderate transverse momenta.
We have checked numerically that this conclusion can be drawn also
for other scale choices ($\mu_0^{\rm (c)},\,\mu_0^{\rm (d)}$),
confirming the almost scale-independent impact of the $\nlotwo$ and
$\nlothree$ corrections.

In \reffi{fig:ptbbtt} we consider the distributions in the transverse
momentum of the $\Pt\bar{\Pt}$ and the $\Pb\bar{\Pb}$ system.
The former is not measurable at the LHC but can be reconstructed
from Monte Carlo truth, choosing the positively-charged lepton--neutrino
pair that best reconstructs the top-quark mass when combined with
the momentum of the bottom quark. The latter variable is directly observable at the LHC.

The $\nloone$ corrections to the $\Pt\bar{\Pt}$ transverse-momentum
distribution [\reffi{fig:pttt}] have already been investigated in \citere{Denner:2020hgg}:
these QCD corrections grow monotonically and become dramatically large and positive
at high $\pt{\Pt\bar{\Pt}}$. Note that at LO this variable coincides with the
$p_{\rm T}$ of the recoiling $\PW^+$~boson, and therefore is sensitive to
the real QCD radiation which is not clustered into $\Pb$ jets (and that
cannot be clustered to the $\PW^+$-boson decay products). The $\nloone$ corrections
receive a sizeable contribution by the $\Pg q/\Pg \bar{q}$ partonic channels,
which are enhanced by the gluon PDF.
In contrast, the $\nlotwo$ ones feature a typical NLO EW behaviour in 
the tail of the distribution, giving a negative and monotonically decreasing
correction to the LO cross-section. Differently from the QCD corrections,
the additional photon can be clustered into any of the external charged particles,
thus also into the decay products of the recoiling $\PW$~boson. 
Furthermore, the cross-section is not
enhanced by the $\gamma q/\gamma \bar{q}$ partonic channels due to the
very small photon luminosity in the proton.
The $\nlothree$ contribution is positive in the whole analyzed spectrum, and increases
from $6\%$ (below 50 GeV) to a maximum of $25\%$ around $\pt{\Pt\bar{\Pt}}=2m_{\Pt}$,
then it slowly decreases in the large-$p_{\rm T}$ region.
Relative to the $\loqcd+\nloone$ result, the combination of all other corrections gives
an effect which is about $10\%$ in the soft-$p_{\rT}$ region and diminishes towards negative
values at large $p_{\rT}$.

The transverse momentum of the $\Pb\bar{\Pb}$ system [\reffi{fig:ptbb}] is 
correlated to the one of $\Pt\bar{\Pt}$ system. The $\nlotwo$ corrections to this observable
behave in the same manner as those for the $\pt{\Pt\bar{\Pt}}$ distributions, giving
a $-20\%$ contribution around $400\GeV$.
The $\nlothree$ corrections grow monotonically from $+10\%$ (at low transverse momentum)
to $+30\%$ (around $400\GeV$).
In the soft part of the spectrum ($\pt{\Pb\bar{\Pb}}<150\GeV$), the $\nloone$ corrections
are rather flat, while in the large-$p_{\rm T}$ region they grow positive and become very large,
similarly to the $\pt{\Pt\bar{\Pt}}$ distribution.
The overall NLO corrections are very small below $150\GeV$ due to mutual cancellations
among the three contributions, while at larger transverse momenta the corrections
are dominated by the $\nloone$ contribution. Furthermore, relatively to the
$\loqcd+\nloone$ distribution, the combined NLO corrections give a flat and positive effect
between $7\%$ and $8\%$ in the whole analyzed spectrum.

In all analyzed transverse-momentum distributions of Figures~\ref{fig:ptet}--\ref{fig:ptbbtt},
the $\loew$ contribution increases monotonically (relatively to the
$\loqcd$ one) but never exceeds $3\%$ of the $\loqcd$ cross-section.

\begin{figure*} 
  \centering
    \subfigure[Invariant mass of the antitop quark. \label{fig:mtbar}]{\includegraphics[scale=0.4]{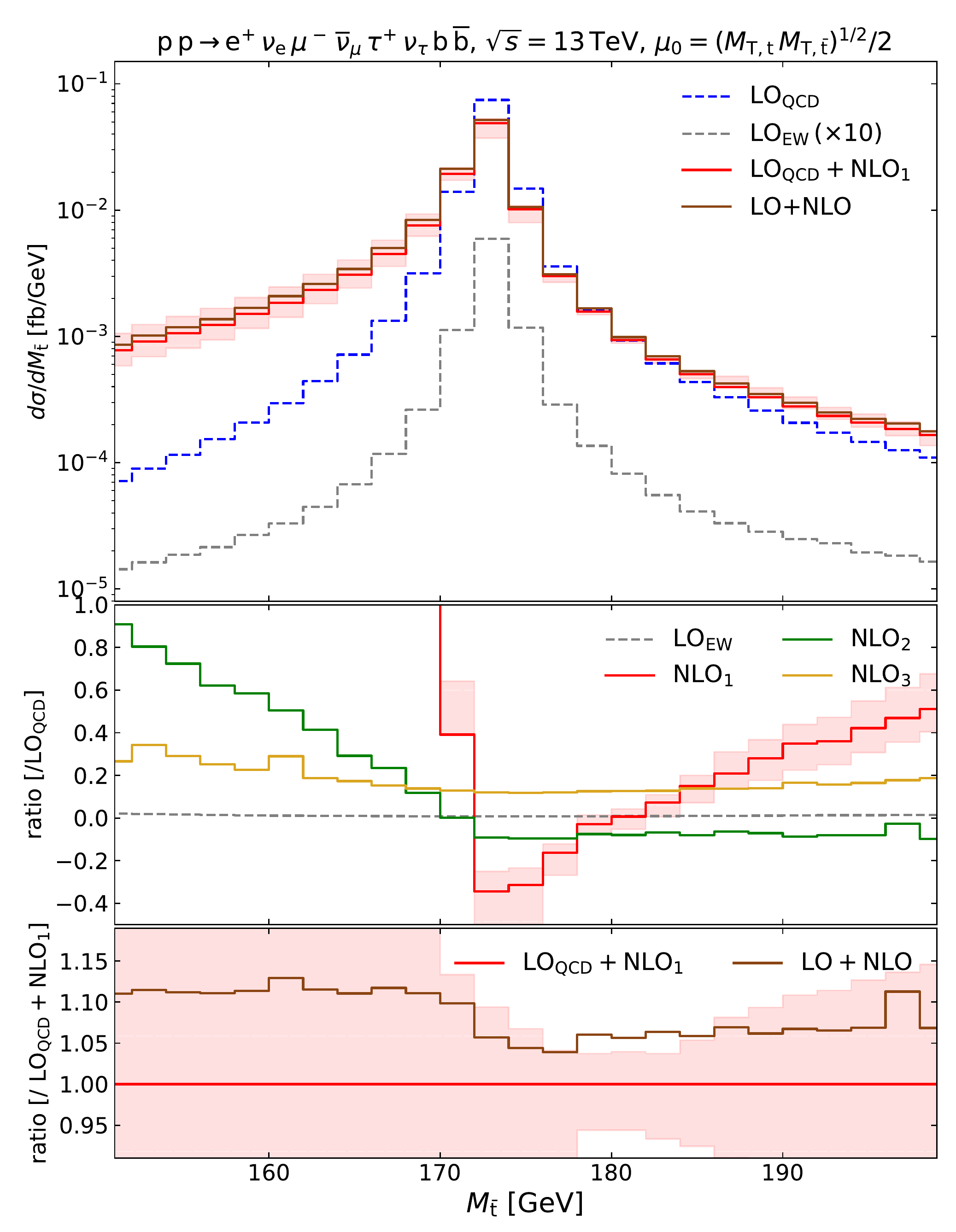}}
    \subfigure[$H_{\rm T}$ variable.\label{fig:ht}]{\includegraphics[scale=0.4]{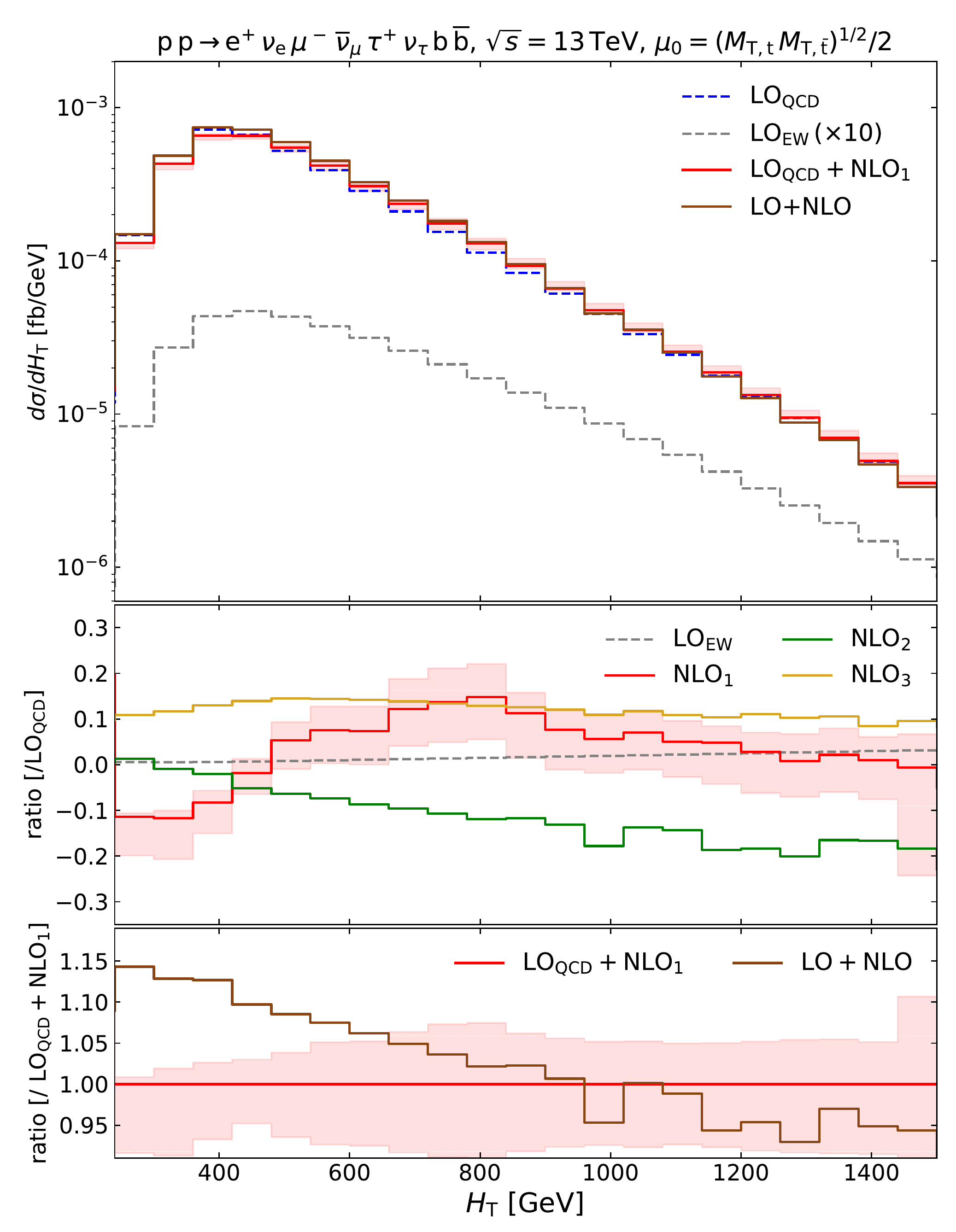} }
  \caption{Distributions in the invariant mass of the antitop quark (left)
    and in the $H_{\rm T}$ variable (right). Same structure as \reffi{fig:ptet}.    
  }\label{fig:mtht}
\end{figure*}

In \reffi{fig:mtbar}, we display the distribution in the invariant mass of
the antitop quark, which in our setup can be reconstructed from the Monte Carlo truth.
The lineshape is dominated by the Breit-Wigner distribution of the
leptonic decay of the antitop quark. The $\nloone$ corrections are negative at the
peak while below the pole mass they give a very large enhancement to the
LO result. Such a radiative tail, coming from unclustered real radiation,
is also present, though less sizeable, in the NLO EW corrections (unclustered photons).
At values larger than the top-quark mass, the distribution receives
an increasingly positive contribution from $\nloone$ corrections, while the $\nlotwo$ ones
give an almost flat correction of $-10\%$  to the $\loqcd$ cross-section. 
The $\loew$ contribution shows a slightly wider distribution than the
$\loqcd$ one. This could be attributed to the relatively larger contribution of non-resonant
background diagrams in the $\loew$ contribution.
Nonetheless, the impact of this difference on the full distribution is
almost invisible owing to the very small size of the $\loew$ contribution.
The $\nlothree$ corrections behave differently from the $\nloone$ ones, giving
 a rather flat enhancement of the fiducial LO cross-section, which is minimal
around the peak ($+11\%$ at $m_{\Pt}$) and mildly increases towards the tails ($+20\%$ at
$200\GeV$, $+30\%$ at $150\GeV$). This is due to the very large contribution of the
$\Pu(\Pc)\Pg$ partonic channel, which has a light $\Pd(\Ps)$ in the final state that cannot
come from the radiative decay of the top or of the antitop quark (differently from
final-state gluons).
As can be seen in the bottom panel of \reffi{fig:mtbar},
the total $\rm LO+NLO$ result is $12\%$ higher than the $\loqcd+\nloone$ one
below the top-quark mass. This enhancement is smaller at ($4\%$)
and above ($7\%$) the top mass.

Another variable that is often investigated in LHC analyses is $H_{\rT}$, whose
definition is given in Eq.~\refeq{eq:htdef}.
The LO and NLO distributions in this observable are shown in \reffi{fig:ht}.
As in the transverse-momentum distributions studied above,
the $\nlotwo$ radiative corrections decrease monotonically towards
large values of $H_{\rT}$ (about $-20\%$ for $H_{\rT}\approx 1.2\TeV$).
The $\loew$ contribution grows to $5\%$ of the $\loqcd$ cross-section at $1.5\TeV$,
where the NLO cross-section is two orders of magnitude lower than its value
at the maximum of the distribution.
The $\nlothree$ corrections are rather flat and enhance
the $\loqcd$ result  between $10\%$ and $15\%$.
The $\nloone$ corrections are characterized by a non-flat shape that is
increasing for $H_{\rT}<800\GeV$ from $-10\%$ to $+25\%$ and
decreasing in the rest of the considered spectrum. 
We further observe that the combination of the three NLO perturbative orders
yields an almost vanishing correction in the soft region of the spectrum,
while in the tail of the distribution the overall correction is dominated by
the $\nlotwo$ contribution for our scale choice. 
In a similar fashion as in other transverse-momentum
distributions, the ratio of the combined $\rm LO+NLO$ result over the
$\loqcd+\nloone$ decreases monotonically from $1.15$ to $0.95$ in the analyzed range. 

\begin{figure*} 
  \centering
    \subfigure[Invariant mass of the ${\Pb\bar{\Pb}}$ system.\label{fig:mbb}]{\includegraphics[scale=0.4]{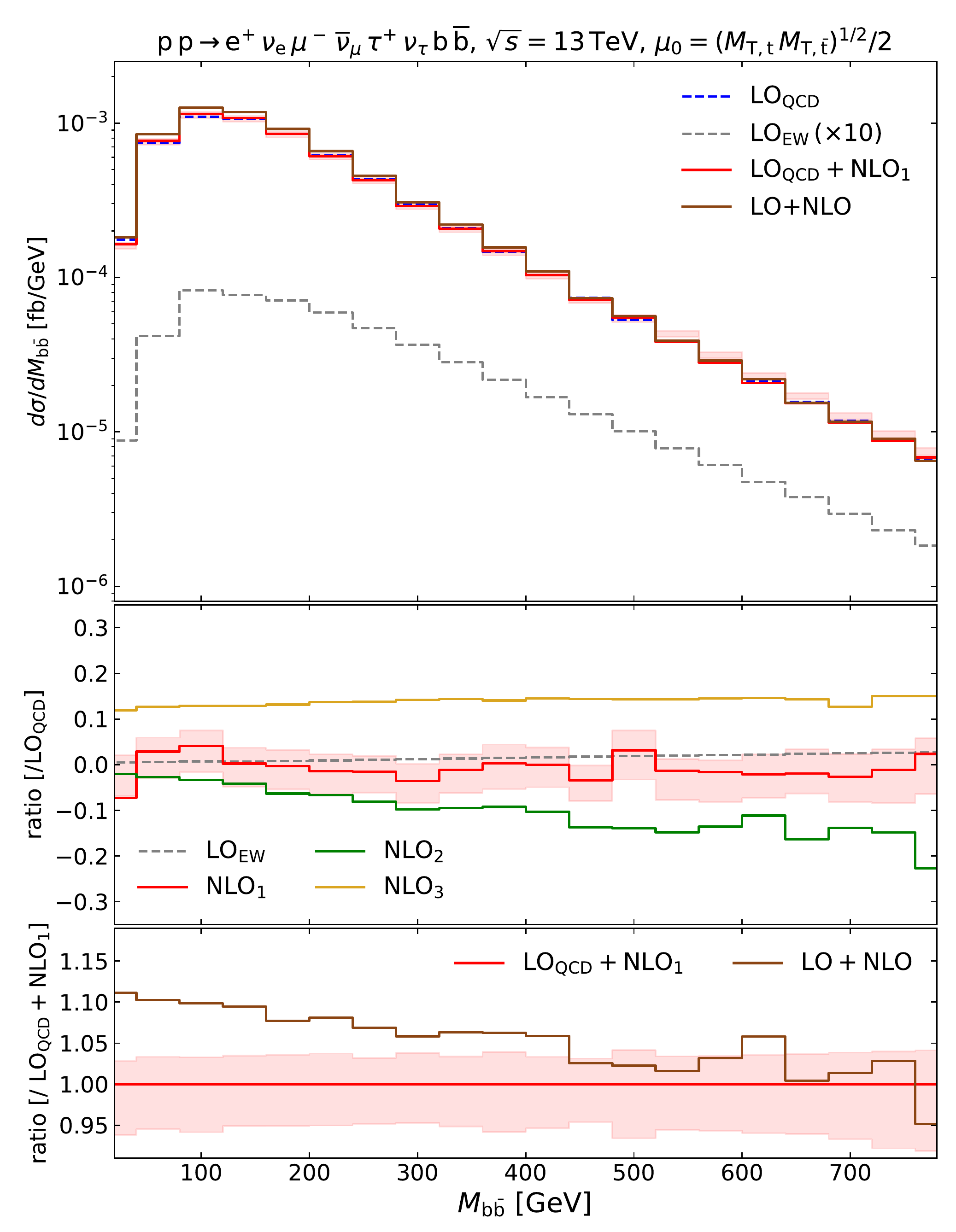}}
    \subfigure[Invariant mass of the three-charged-lepton system.\label{fig:m3l}]{\includegraphics[scale=0.4]{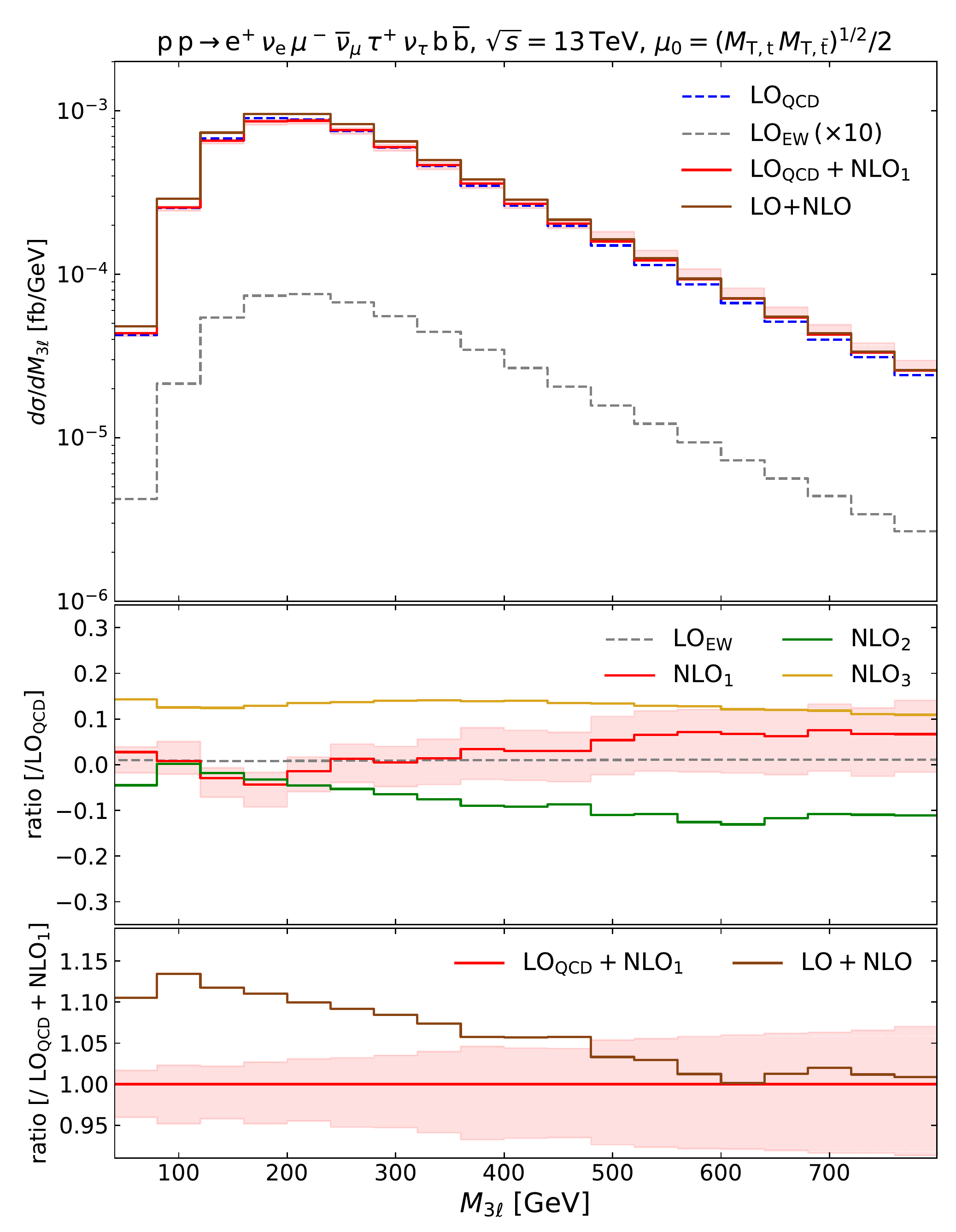} }
  \caption{Distributions in the invariant mass of the $\Pb\bar{\Pb}$ system (left)
    and of the three-charged-lepton system (right). Same structure as \reffi{fig:ptet}.}\label{fig:mbb3l}
\end{figure*}

In \reffi{fig:mbb3l} we study more invariant-mass distributions.
%
The distribution in the invariant mass of the two-b-jet system
[\reffi{fig:mbb}] is characterized by rather flat QCD corrections ($\nloone$ and $\nlothree$).
The $\nlothree$ corrections enhance  the $\loqcd$ cross-section  by
$11\%$ to $14\%$ everywhere in the analyzed invariant-mass range. 
The $\nlotwo$ contribution has a similar behaviour as the one found for the previous variables,
growing negative towards the tail of the distribution.

The distribution in the invariant mass of the three-charged-lepton system
is considered in \reffi{fig:m3l}. 
The behaviour of the $\nlotwo$ and $\nlothree$ corrections follows closely the one for the
$\Pb\bar{\Pb}$ system, apart from a less steep decrease of the EW corrections towards
large invariant masses. These corrections are at the $-10\%$ level for
masses larger than $500\GeV$.
The $\nloone$ corrections vary by hardly more than $10\%$  in the studied range.

As shown in the bottom panels of \reffi{fig:mbb3l}, both for the $\Pb\bar{\Pb}$ system
and for the three-charged-lepton system, the inclusion of $\nlotwo$ and $\nlothree$
corrections (as well as of $\loew$, though hardly visible) gives a non-flat correction
to the NLO QCD invariant-mass distributions, decreasing monotonically from $+12\%$ to zero in the
considered spectra.

\begin{figure*} 
  \centering
    \subfigure[Rapidity of the muon.\label{fig:ymu}]{\includegraphics[scale=0.4]{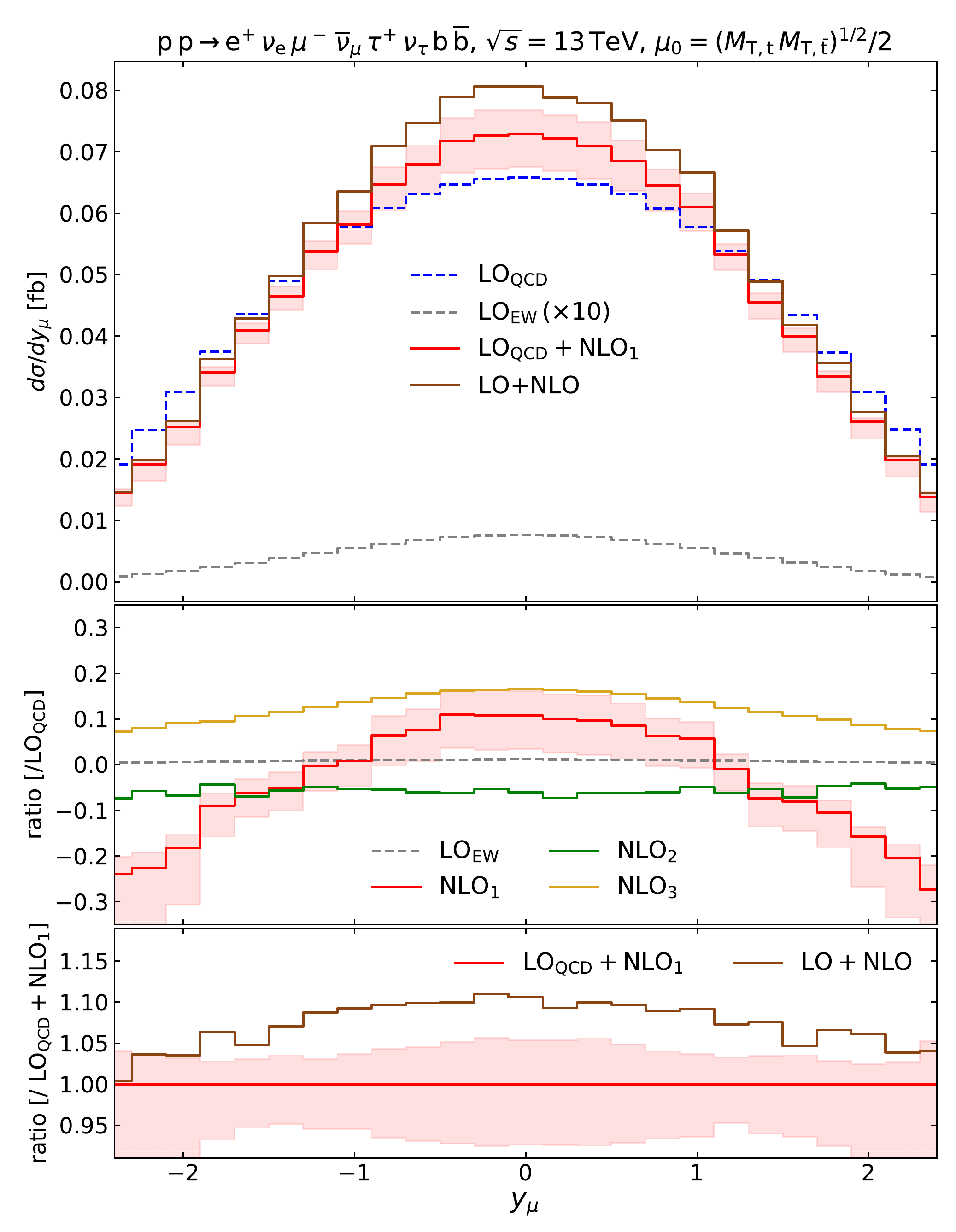} }
    \subfigure[Rapidity of the antitop quark.\label{fig:ytbar}]{\includegraphics[scale=0.4]{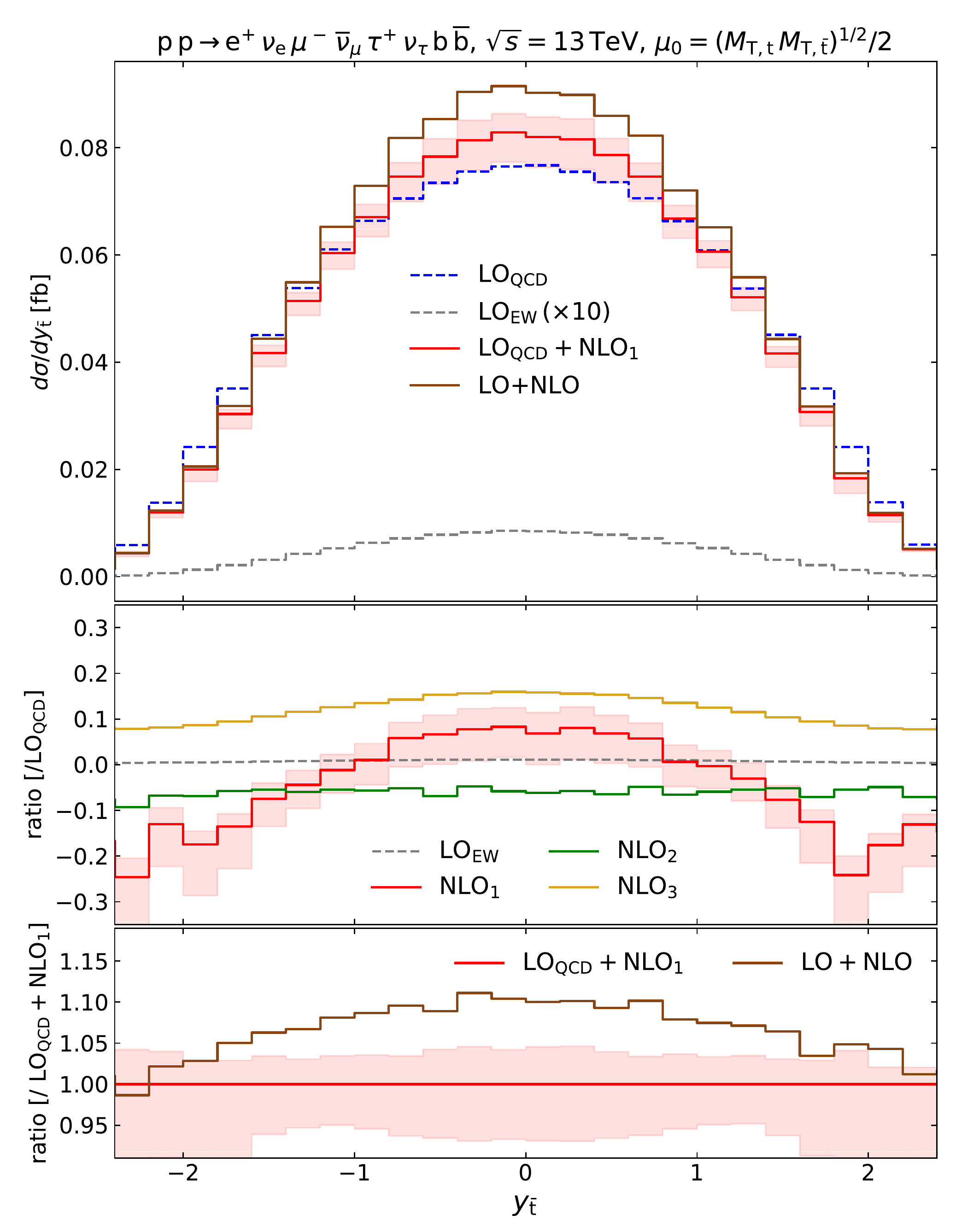} }
  \caption{Distributions in the rapidity of the muon (left)
    and of the antitop quark (right). Same structure as \reffi{fig:ptet}.}\label{fig:rap}
\end{figure*}

After presenting transverse-momentum and invariant-mass distributions,
we switch to some relevant angular variables.
In \reffi{fig:ymu} and \reffi{fig:ytbar} we display the distributions in
the rapidity of the muon and of the antitop quark, respectively.
%
\begin{figure*} 
  \centering
    \subfigure[Azimuthal angle between the positron and the muon.\label{fig:azi}]{\includegraphics[scale=0.4]{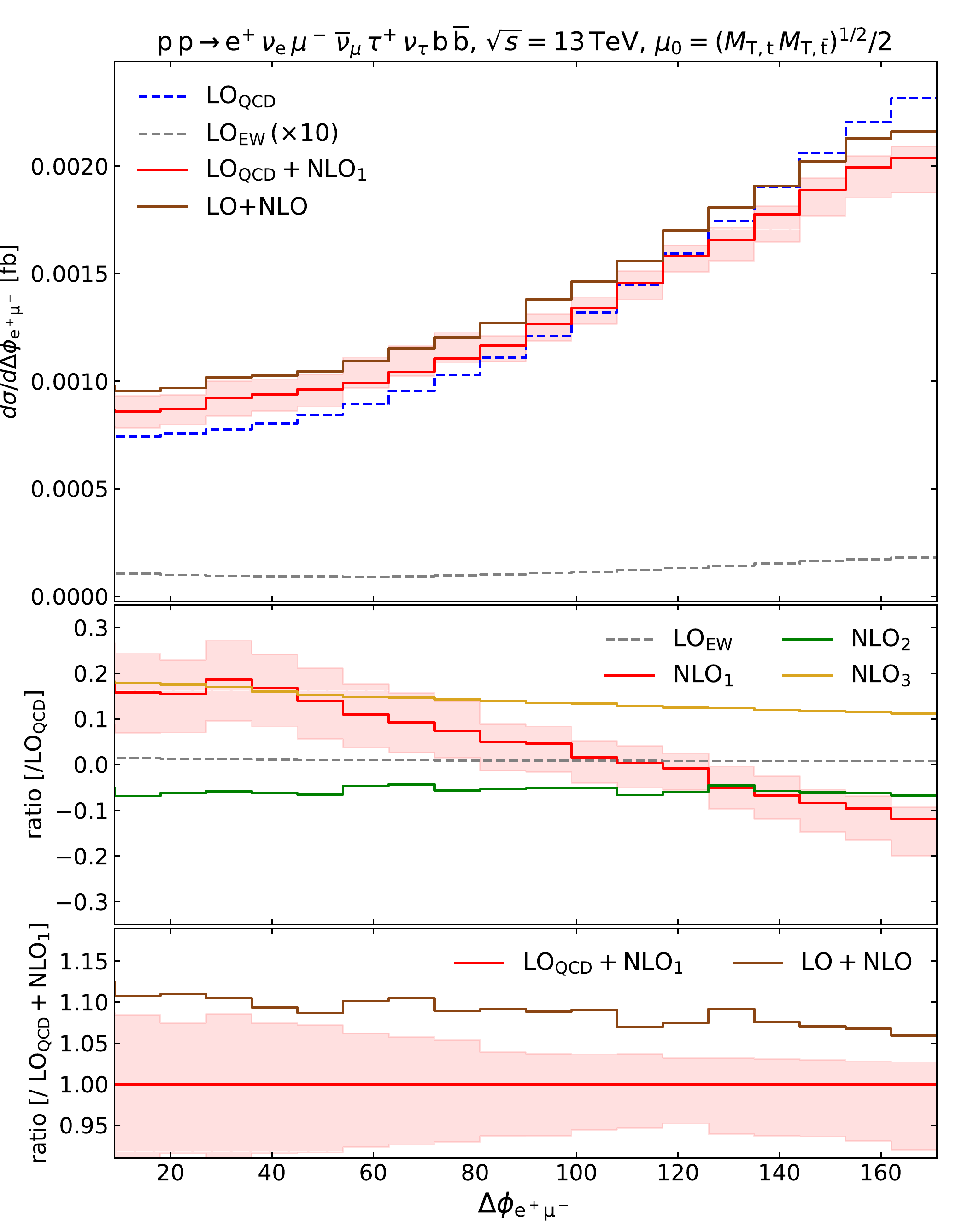}}
    \subfigure[$R$ distance between the two b jets.\label{fig:dist}]{\includegraphics[scale=0.4]{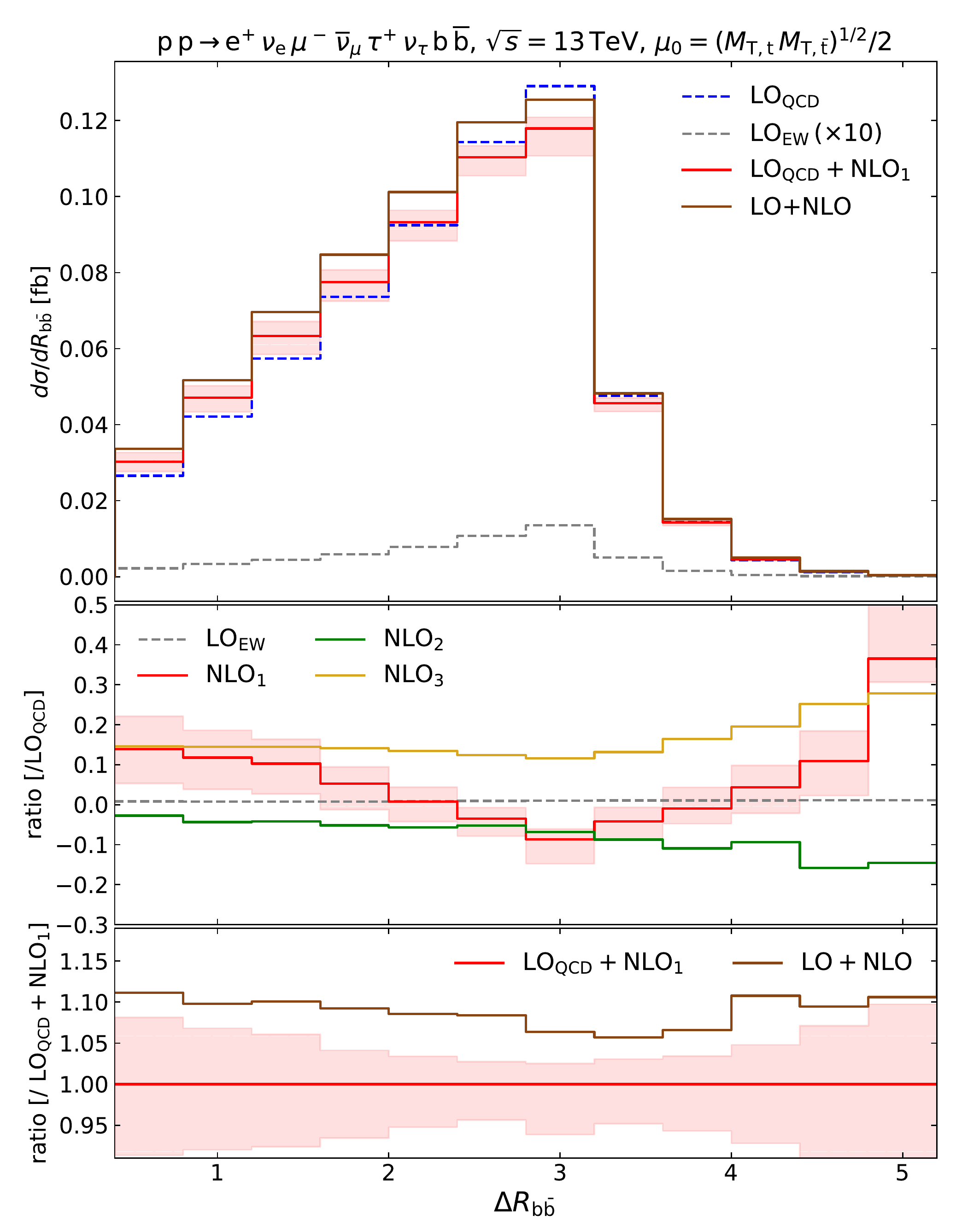}} 
  \caption{Distributions in the azimuthal difference between the positron and the muon (left)
    and in the azimuthal-angle--rapidity distance between the $\Pb$ jet and the $\bar{\Pb}$ jet (right). Same structure as \reffi{fig:ptet}.}\label{fig:azidist}
\end{figure*}
%
Since these two variables
are correlated (the dominant resonant structure involves the decay
$\bar{\Pt}\rightarrow \bar{\Pb}\mu^-\bar{\nu}_\mu$),
the muon rapidity, which is precisely measurable
at the LHC, represents a suitable proxy for the rapidity of the
corresponding antitop quark (which can only be reconstructed from
Monte Carlo truth).
Note that the muon rapidity is sharply cut at $\pm 2.5$ by fiducial
selections, while this is not the case for the antitop quark. However,
thanks to the rapidity cut applied to b jets, the cross-section
is strongly suppressed for $|y_{\bar{\Pt}}|>2.5$. Both the muon and
the antitop quark are produced preferably in the central region.
The $\nlotwo$ corrections
are rather flat, giving between $-4\%$ and $-8\%$ decrease to the LO
cross-section.
The relative $\nloone$ corrections to the muon-rapidity distribution
are characterized by a large variation  (about $35\%$) in the available
range. 
Relative to $\loqcd$, the differential $\nlothree$ corrections
have a similar shape as the $\nloone$ ones, giving in the whole rapidity
range a positive correction ($8\%$ in the forward regions, $16\%$ in the central region).
Almost identical results are found in the rapidity distribution of the antitop~quark.
Owing to the $\nlothree$ corrections, the ratio between the complete  NLO
prediction and  the $\loqcd+\nloone$ one has a maximum of $1.11$ in the central region and
diminishes towards forward regions (close to unity). This holds both
for the muon and for the antitop-quark rapidity spectra.

In \reffi{fig:azi} we consider the distribution in the azimuthal separation
between the positron and the muon. 
The two charged leptons tend to be produced in opposite directions
both at LO and at NLO, but the inclusion of radiative corrections enhances
the fraction of events with small azimuthal separations.
The $\nlotwo$ corrections are negative and roughly constant 
($-5\%$ to $-7\%$) over the full angular range,
while the $\nlothree$ contribution decreases monotonically from $+18\%$ to $+11\%$
relatively to the $\loqcd$ result.
As already observed in \citere{Denner:2020hgg}, the $\nloone$ correction to the LO QCD
cross-section decreases with an almost constant negative slope over the full range.
The overall NLO corrections to the $\loqcd$ cross-section are positive
everywhere except in the vicinity of the peak at $\Delta\phi_{\Pe^+\mu^-}=\pi$.
Relative to the $\loqcd+\nloone$ prediction, the combination of $\nlotwo$ and
$\nlothree$ corrections gives a pretty flat enhancement ($1.11$ at $\Delta\phi_{\Pe^+\mu^-}=0$,
$1.06$ at $\Delta\phi_{\Pe^+\mu^-}=\pi$).

As a last differential result, we present in \reffi{fig:dist} the distribution
in the $R$~distance between the two b~jets [see Eq.~\refeq{eq:Rdist} for its definition].
This distribution is characterized by an absolute maximum around $\Delta R_{\Pb\bar{\Pb}}\approx\pi$.
The negative $\nlotwo$ corrections diminish monotonically over the analyzed spectrum.
At large distance ($\Delta R_{\Pb\bar{\Pb}}>5$) they give a contribution of $-15\%$.
The positive $\nlothree$ corrections diminish from $+15\%$ at $\Delta R_{\Pb\bar{\Pb}}\approx 0$
to $+12\%$ at $\Delta R_{\Pb\bar{\Pb}}\approx \pi$ and then  increase again.
The $\nloone$ ones show a similar behaviour, however, with larger slopes.
The combined $\nlotwo$ and $\nlothree$ corrections enhance the
$\loqcd+\nloone$ prediction between $6\%$ and $11\%$, similarly to the case
of the azimuthal distance shown in \reffi{fig:azi}, but with a somewhat
different shape.

The results for the differential distributions presented
in \reffis{fig:ptet}--\ref{fig:azidist} show that in many kinematic
regions the $\nlotwo$ and $\nlothree$ corrections give an
enhancement at the level of $10\%$ to the $\loqcd+\nloone$ result
that is larger than the QCD scale uncertainties at the same
perturbative order. This concerns in particular the soft- and
moderate-$p_{\rm T}$, the low-mass and the central-rapidity regions,
which are also the statistically most-populated ones. This reinforces
that including formally subleading corrections ($\nlotwo$,
$\nlothree$) is necessary to give a more realistic description of
total and differential $\Pt\bar{\Pt}\PW$ cross-sections.  

\section{Conclusions}\label{concl}
In this work we have presented the NLO corrections to the off-shell production
of $\Pt\overline{\Pt}\PW^+$ at the LHC in the three-charged-lepton channel.
These include the next-to-leading-order (NLO) QCD corrections
to the QCD ($\nloone$) and to the electroweak leading order ($\nlothree$), as well as
the NLO electroweak corrections to the QCD leading order ($\nlotwo$).
It is the first time that the $\nlotwo$ and $\nlothree$ radiative corrections are computed
with full off-shell dependence for a physical final state, accounting for all
non-resonant, interference, and spin-correlation effects.

Both integrated and differential cross-sections have been presented and
discussed in a realistic fiducial region, keeping in mind the limited statistics
of the LHC data and relating the off-shell description of the process to the
inclusive predictions that are available in the literature.

The $\nlotwo$ and $\nlothree$ corrections give a $-5.5\%$ and
a $+13\%$ contribution, respectively, to the LO cross-section, almost independently of
the choice of the factorization and renormalization scales.
%
The sizeable impact of $\nlotwo$ and $\nlothree$ corrections
makes it essential to combine them with the $\nloone$ ones, in order to
arrive at reliable predictions.

The theory uncertainties from 7-point scale variations are
driven by the $\nloone$ corrections, which are the only corrections which
feature a NLO-like scale dependence. Their inclusion reduces the scale
uncertainty of the LO cross-section from $20\%$ to $5\%$.

The investigation of differential distributions reveals a more
involved interplay among the various perturbative orders compared
with the integrated results.
The $\nlotwo$ corrections drop by up to $-20\%$
in most of the transverse-momentum and invariant-mass distributions,
showing the typical behaviour of
EW corrections with large Sudakov logarithms at high energies.
They are rather flat for angular observables.
The $\nlothree$ corrections give a positive enhancement between $+10\%$ and $+20\%$
($30\%$ in some cases) to the LO cross-section in all analyzed distributions,
They are dominated by the $\Pu\Pg$ partonic channel
(formally belonging to QCD real corrections to LO EW)
that embeds $\Pt\PW$ scattering.
The $\nloone$ corrections, which have already been presented in the literature,
show quite variable patterns in the various
differential $K$-factors.

We stress that all three NLO contributions usually give non-flat corrections
to the LO distributions, also to the angular ones. This indicates
that rescaling QCD results (either LO or NLO accurate) by flat $K$-factors
could result in a bad description of some LHC observables.

In the light of an improved experimental description of the
$\Pt\overline{\Pt}\PW$ process,
the inclusion of decay and off-shell effects is mandatory.
Although for sufficiently inclusive observables the full computation is well
approximated by on-shell calculations, the inclusion of off-shell effects
in the modeling of $\Pt\overline{\Pt}\PW$ production is definitely needed when
studying the tails of transverse-momentum and invariant-mass
observables.

\section*{Acknowledgements}
We thank Mathieu Pellen for useful discussions on photon-induced
electroweak real corrections, Timo Schmidt for performing checks with
\mocanlo, and Jean-Nicolas Lang and Sandro Uccirati for maintaining
\recola.  This work is supported by the German Federal Ministry for
Education and Research (BMBF) under contract no.~05H18WWCA1.

\bibliographystyle{elsarticle-num}
\bibliography{ttv}

\end{document}